\documentclass[rmp,notitlepage,nolongbibliography,amssymb]{revtex4-1}
\usepackage{graphicx}
\usepackage{amsthm}
\usepackage{amssymb}
\usepackage{color}
\usepackage[normalem]{ulem}
\usepackage{multirow}
\def \A{ {\cal A}}
\def \proclaim#1#2{\medskip\noindent{\bf #1\enskip}{\sl #2}}
\def \pf{\medskip\noindent{\bf Proof.\enskip}}
\def \qed{\par\rightline{$\square$}\par\medskip\noindent}
\def\nnmo{$(N\!-\!1,1)$}

\def\Tnmo{$T^{N\!-\!1}$}
\setcitestyle{numbers,square}
\begin{document}
\def\EEE{\end{equation}}
\def\etal{{\it et. al.}}
\def\b{\bigskip\noindent}
\newcommand{\mod}{\mathrel{{\rm mod}}}
\def\Re{\mathop{\rm Re}\nolimits}
\def\Im{\mathop{\rm Im}\nolimits}
\def\OPF{\Omega_{op}}
\title{Dynamics of the Kuramoto-Sakaguchi Oscillator Network
with Asymmetric Order Parameter}
\author{Bolun Chen$^{1}$, Jan R.~Engelbrecht$^{1}$ and Renato Mirollo$^{2}$}
\address{Departments of Physics$^{1}$ and \mbox{Mathematics,$^{2}$}
 Boston College, Chestnut Hill, MA 02467}
\begin{abstract}
{\bf Abstract:}
We study the dynamics of a generalized version of the famous Kuramoto-Sakaguchi coupled oscillator model.  In the classic version of this system, all oscillators are governed by the same ODE, which depends on the order parameter of the oscillator configuration.  The order parameter is the arithmetic mean of the configuration of complex oscillator phases, multiplied by some constant complex coupling factor.  In the generalized model we consider, the order parameter is allowed to be any complex linear combination of the complex oscillator phases, so the oscillators are no longer necessarily weighted identically in the order parameter.  This asymmetric version of the K-S model exhibits a much richer variety of steady-state dynamical behavior than the classic symmetric version; in addition to stable synchronized states,  the system may possess multiple  stable (N-1,1) states, in which all but one of the oscillators are in sync, as well as multiple families of neutrally stable asynchronous states or closed orbits, in which no two oscillators are in sync.  We present an exhaustive description of the possible steady state dynamical behaviors; our classification depends on the complex coefficients that determine the order parameter.  We use techniques from group theory and hyperbolic geometry to reduce the dynamic analysis to a 2D flow on the unit disc, which has geometric significance relative to the hyperbolic metric.  The geometric-analytic techniques we develop can in turn be applied to study even more general versions of Kuramoto oscillator networks.
\end{abstract}
\pacs{05.45.Xt,74.81.Fa}
\maketitle

\noindent
The Kuramoto-Sakaguchi coupled oscillator model is a famous, well-studied
dynamical system that models the dynamics of a network of $N$ identical coupled
oscillators.  In the classic formulation, all the oscillators are driven by
identical coupling to the system's order parameter, which is just the average
of the complex oscillator phases, multiplied by some complex coupling constant.
We study a generalized version of the Kuramoto-Sakaguchi system, in which the
order parameter is now a complex-linear combination of the complex oscillator
phases, but not necessarily with identical coefficients.  We analyzed the
dynamics of this asymmetric K-S system, and found that it supports a much
richer variety of dynamical behaviors than the classic K-S model, which
typically has steady state dynamics that are completely synchronized or
completely asynchronous with all oscillators out of sync with each other.  The
asymmetric K-S model also can support multiple stable \nnmo\ states, in which
all but one of the oscillators are in sync, as well as multiple families of
neutrally stable asynchronous states or closed orbits.  We introduce new
group-theoretic and geometric techniques to study the asymmetric K-S model,
which effectively reduce the $N$-dimensional dynamics to a 2D flow on the unit
disc, and use the natural hyperbolic geometry on the disc to study this flow.
The techniques we develop lead to a complete classification of the dynamics of
the asymmetric K-S model in terms of the order parameter coefficients, and can
be used to study more general oscillator networks via a similar
dimensional reduction.  This also builds a connection between the somewhat
distant fields of oscillator network dynamics and hyperbolic geometry / low-D
complex dynamics.  

\section{Introduction}

Our  subject is the study of networks of Kuramoto oscillators, which are dynamical systems governed by equations of the form
\begin{equation}\label{Ksys}
 \dot \theta_j = A + B \cos \theta_j + C \sin \theta_j, \quad j = 1, \dots, N. 
\end{equation}
Here $\theta_j$ is an angular variable (i.e.~an element of ${\Bbb R}  \mod 2\pi \Bbb Z$) and the coefficients $A, B, C$ are smooth functions of $(\theta_1, \dots, \theta_N)$.  The state space for this system is the $N$-fold torus $T^N = (S^1)^N$.   Kuramoto oscillator networks often arise as idealized models of physical dynamical systems, like Josephson junction series 
arrays~\cite{nichols1992ubiquitous,strogatz1993splay}, 
and also as the result of averaging more complex dynamical 
systems~\cite{swift1992averaging}.  
Beginning with the original work of Kuramoto over forty years 
ago~\cite{kuramoto1975self}, 
Kuramoto networks have been a very fertile research subject in applied 
dynamics~\cite{pikovsky2015dynamics}.  
When the functions $A, B, C$ are symmetric in the variables $\theta_j$, we say the system is a network of {\sl identical} Kuramoto oscillators; any permutation of the components of a solution $(\theta_1(t), \dots, \theta_N(t))$ results in another solution to the system.   This is the case  for the famous Kuramoto-Sakaguchi (K-S) model, which has equations
\begin{equation}\label{Kphase}
\begin{array}{rcl}
 \dot \theta_j &=& 
\displaystyle \omega + {K \over N} \sum_{k = 1}^N  \sin (\theta_k - \theta_j + \psi) 
\\ &=& 
\displaystyle \omega + \left( {K \over N} \sum_{k = 1}^N  \sin (\theta_k  + \psi) \right) \cos \theta_j -  \left( {K \over N} \sum_{k = 1}^N  \cos (\theta_k  + \psi) \right) \sin \theta_j, 
\quad
\quad j = 1, \dots, N; \ \omega, K, \psi \ {\rm constants}.
\end{array}
\end{equation}

In this paper we will investigate the dynamics of Kuramoto networks where the functions $A, B, C$ are {\sl not} symmetric in the $\theta_j$.  As we shall see, dropping the symmetry assumption leads to a richer variety of dynamic behavior.  The focus of our work is a variation of the K-S model: the asymmetric K-S network given by
\begin{equation}\label{AKS}
\begin{array}{rcl}
 \dot \theta_j &=& 
\displaystyle \omega + \sum_{k = 1}^N  r_k\sin (\theta_k - \theta_j + \psi_k) 
\\ &=& 
\displaystyle \omega +\left(  \sum_{k = 1}^N  r_k\sin (\theta_k  + \psi_k) \right) \cos \theta_j -\left(  \sum_{k = 1}^N  r_k\cos (\theta_k  + \psi_k) \right) \sin \theta_j   , 
\quad
\quad j = 1, \dots, N; \  \omega, r_k, \psi_k \ {\rm constants}.
\end{array}
\end{equation}
The dynamics of this network are governed by its {\sl order parameter}, which we can express in complex form, with $c_j = a_j + i b_j = r_j e^{i \psi_j}$  and $z_j = e^{i\theta_j}$, as
\begin{equation}\label{AOP}
\A = \sum_{j = 1}^N c_j z_j. 
\end{equation}
It turns out, broadly speaking, that the dynamics depend largely on the sum of the $c_j$, which we denote by
$$
c =  a + ib = \sum_{j=1}^N c_j.
$$
As we discussed 
in~\cite{chen2017hyperbolic},
the system (\ref{AKS}) is invariant under the phase shift $\theta_j(t) \mapsto \theta_j(t) + \theta_0$ for any constant $\theta_0$.  Hence we can identify states which are equal up to a phase shift, and reduce the dynamics to an \nnmo-dimensional state space, which is the torus $T^{N-1}$.  In this reduced state space, there is a {\sl unique} state with all $\theta_j$ equal, which we refer to as the {\sl sync state} or just {\sl sync}.

Our goal is to understand the generic long-term behavior of trajectories in the reduced state space, in forward and backward time.  
Some terminology: an {\it asynchronous state} has all $\theta_j$ distinct;
an \nnmo\ {\it state} has all but one $\theta_j$ equal.
The first result, Theorem 1, is that if $a > 0$ then almost all trajectories in the reduced state space converge in forward time to sync, and in backward time to an asynchronous state or to one of finitely many \nnmo\ states.
The dynamics are similar if $a<0$, except reversed in time.  
In the course of preparing this manuscript we learned that this result was independently discovered by M.~A.~Lohe
in~\cite{lohe2017ws},
Lohe presents an argument that is essentially correct,
but has some subtle technical gaps 
which we address in the discussion following the proof of Theorem 1.
Our approach, which is different than 
in~\cite{lohe2017ws},
is based on the 
correspondence between the system dynamics and a flow on the hyperbolic disc with special attention to the
behavior at the boundary circle.
The geometric techniques we develop in preparation for the proof of Theorem~1 
also form the basis of the proofs of our subsequent Theorems~2-5.

The case $a = 0$ is covered in Theorem 2; then the system has Hamiltonian structure, and almost all trajectories 
in the reduced state space are periodic or homoclinic connections to and from sync, with one exceptional case, which is if 
at least one of the coefficients satisfies 
the condition $b_j = b/2$.   Then in addition to the behavior above, there also exists a positive measure set of initial conditions with trajectories that converge in forward time to \nnmo\ states (and also a positive measure set of initial conditions  with trajectories that converge in backward time to \nnmo\ states).  We will also discuss three special cases of this system, where we can describe additional details of the dynamics:  the case $c = 0$, where the dynamics have both  gradient and Hamiltonian structure; 
the case of real $c_j$ with $c = 0$, where the dynamics 
flow along a 2D electrostatic field,
and the case of real $c_j > 0$, which includes the classic Kuramoto model with all $c_j = K/N$.  The results are summarized in Table 1 in the Discussion section at the end of this paper.

The organization of this paper is as follows: We begin by summarizing some of our earlier work on Kuramoto networks 
in~\cite{chen2017hyperbolic},
which exploits a connection to hyperbolic geometry to simplify the analysis of the network dynamics.  We will then derive some general properties of the system (\ref{AKS}), and prove Theorems 1 and 2.  After this we proceed with the analysis of the special cases of (\ref{AKS})\ mentioned above, and conclude with some suggestions for future research.

\section{Complex Formulation}

It is desirable to express the system (\ref{Ksys}) in complex form, with $z_j = e^{i \theta_j}$.  Let $\A = -C + i B$; $\cal A$ is a complex-valued function on $T^N$ which we define to be the {\sl order parameter} for the system.  Then using $\dot z_j = i z_j \dot \theta_j$ we obtain governing equations
\begin{equation}\label{KsysC}
\dot z_j =i A z_j +i z_j {\rm Im} ( \A \overline z_j ) =  i A z_j + {1 \over 2} \left( \A - \overline \A z_j^2 \right), \quad j = 1, \dots, N. 
\end{equation}
The asymmetric Kuramoto-Sakaguchi system (\ref{AKS}) has $A = \omega$ and order parameter $\A$ given by (\ref{AOP}) above.
We will assume henceforth that 
the function $A=\omega$ and that
the order parameter has this form.

\section{Reduction To 3D System}

In their seminal 
paper~\cite{watanabe1994constants}, Watanabe and Strogatz demonstrated that the trajectories for any system of the form (\ref{Ksys}) are constrained to lie in submanifolds of the state space $T^N$ with dimension at most three.  Subsequently, it was shown that these submanifolds are the group orbits under a natural action of the M\"obius group $G$ on the 
torus~\cite{marvel2009identical}.  
Here $G$ is the 3D group of M\"obius transformations which preserve the unit disc $\Delta$ (and hence its boundary $S^1$).   An element $M \in G$ can be expressed uniquely in the form
\begin{equation}\label{Mob}
M(z) = \zeta {z - w \over 1 - \overline w z}, \quad z \in \Bbb C, 
\end{equation}
where the parameters $w$ and $\zeta$ satisfy $|w| < 1$ and $|\zeta| = 1$.   When $\zeta = 1$, we denote the above M\"obius transformation by $M_w$.  If $M \in G$ and $p = (z_1, \dots, z_N) \in T^N$ then
$$
Mp = (M(z_1), \dots, M(z_N))
$$
defines the group action of $G$ on $T^N$.  The group orbits are the sets $Gp = \{Mp \ | \ M \in G\}$.

Now fix a point $p = (\beta_1, \dots, \beta_N)  \in T^N$, and assume that at least three of the $\beta_j$ are distinct.  Then any point $(z_1, \dots, z_N)$ in the group orbit $Gp$ can be expressed in the form $\zeta M_w p$ for a {\sl unique choice} of $w \in \Delta$ and $\zeta \in S^1$.  In effect, $w$ and $\zeta$ can be thought of as coordinates on the group orbit $Gp$.  As derived 
in~\cite{chen2017hyperbolic},
the system (\ref{KsysC}) on $Gp$ is equivalent to the system in $w$ and $\zeta$ given by
\begin{equation}\label{Eqwzeta}
\begin{array}{rcl}
\dot w &=&\displaystyle  -{1 \over 2} ( 1 - |w|^2 )\overline{\zeta} \A(\zeta M_wp) \\
\dot \zeta &=&\displaystyle iA(\zeta M_wp) \zeta - {1 \over 2} \left( \overline w\A(\zeta M_wp) - w \overline{ \A(\zeta M_wp) }\zeta ^2  \right).
\end{array}
\end{equation}

\section{Reduction To 2D System}

When the order parameter function $\cal A$ has the form  (\ref{AOP}), we can cancel the $\zeta$ and $\overline \zeta$ in the $\dot w$ equation above, which then simplifies to an equation in $w$ alone:
\begin{equation}\label{weq}
\dot w =  -{1 \over 2} ( 1 - |w|^2 )\A(M_wp) = -{1 \over 2} ( 1 - |w|^2 ) \sum_{j = 1}^N c_j M_w (\beta_j).  
\end{equation}
The $w$ variable determines a point on the group orbit $Gp$ up to rotation by some $\zeta \in S^1$; in effect, $w$ determines the {\sl phase relations} among the coordinates $z_1, \dots, z_N$.  

More formally, if we identify $p$ and $\zeta p$ for any $\zeta \in S^1$, then the full state space for this reduced model is the $(N\!-\!1)$-dimensional torus $T^{N-1}$; the group orbits $Gp$ under this identification give us reduced group orbits $\widetilde{Gp}$, which are invariant under the reduced dynamics.  In this reduced state space, {\sl sync} is the {\sl unique} fully synchronized state  represented by any $p = (\beta, \dots, \beta)$.  For a base point $p$ with at least three distinct coordinates, its reduced $G$-orbit is parametrized by $w \in \Delta$, and equation (\ref{weq}) gives the dynamics on the reduced orbit.  Fixed points in the reduced system correspond to either fixed points or uniformly rotating solutions with constant phases in the original $N$-dimensional system.  Since we are primarily interested in how the phases among the coordinates evolve, we will henceforth make this reduction and work with the 2D dynamical system given by (\ref{weq}).

Observe that changing the signs of all the $c_j$ in \ref{weq}) is equivalent to reversing the direction of time for the system; this time-reversal property will be used frequently in the sequel.

\section{Boundary Correspondence}

Fix a base point $p = (\beta_1, \dots, \beta_N)$ with all $\beta_j$ distinct; then $w \in \Delta$ parametrizes all possible phase configurations in the reduced group orbit $\widetilde{Gp}$.  We wish to describe what happens to these phase configurations as $w$ approaches the boundary of the disc.  Suppose a sequence $w_n \in \Delta$ converges to some $\beta \in S^1$, and $\beta \ne \beta_j$ for all $j$.  Then
$$
\lim_{n \to \infty} M_{w_n} p = \left( {\beta_1 - \beta \over 1 - \overline \beta \beta_1}, \dots, {\beta_N - \beta \over 1 - \overline \beta \beta_N}\right) = -\beta(1, \dots, 1),
$$
which corresponds to the sync state in the reduced state space $T^{N-1}$.
However, if $w_n \to \beta_j$ for some $j$, then $\lim_{n \to \infty} M_{w_n} \beta_j$ need not exist.  To see this, suppose for example $w_n \to \beta_1$. 
Write $w_n = (1 + i r_n e^{i \theta_n})\beta_1$, with $r_n > 0$, $0 < \theta_n < \pi$, so $\theta_n$ is the angle at which $w_n$ is approaching the boundary circle.  Then
$$
M_{w_n} (\beta_1) = { -  i r_n e^{i \theta_n} \beta_1\over  i r_n e^{-i\theta_n}} = - e^{2i \theta_n}\beta_1.
$$
Therefore $\lim_{n \to \infty} M_{w_n} p $ exists iff $\lim_{n \to \infty} \theta_n = \theta$ exists, and in this case
$$
\lim_{n \to \infty} M_{w_n}p =  -\beta_1(e^{2i \theta}, 1, \dots, 1).
$$
Thus we see that if $w_n \to \beta_j$ for some $j$, and the approach angle  is asymptotically $\theta$, then the limiting configuration of $M_{w_n} p$ in the reduced state space $T^{N-1}$  is the \nnmo\ state in which the $j$th oscillator has phase $2\theta$ relative to the 
others in sync.

If all the coordinates $\beta_j$ are distinct, then this analysis shows that the boundary of the reduced $G$-orbit $\widetilde{Gp}$ consists of all \nnmo\ states together with the sync state.  This is topologically a union of $N$ circles, all meeting at the sync state.  (If $p$  does not have all distinct $\beta_j$, then the boundary of $\widetilde{Gp}$ will consist of $r$ circles meeting at sync, where $r$ is the number of distinct $\beta_j$, provided $r \ge 3$.  If $r =2$ then the boundary is sync.) 
In the next section we describe the dynamics on these boundary components, 
which are invariant for any system with order parameter given by (\ref{AOP}).

\begin{figure}[h]
\centerline{\includegraphics[width=4.4in]{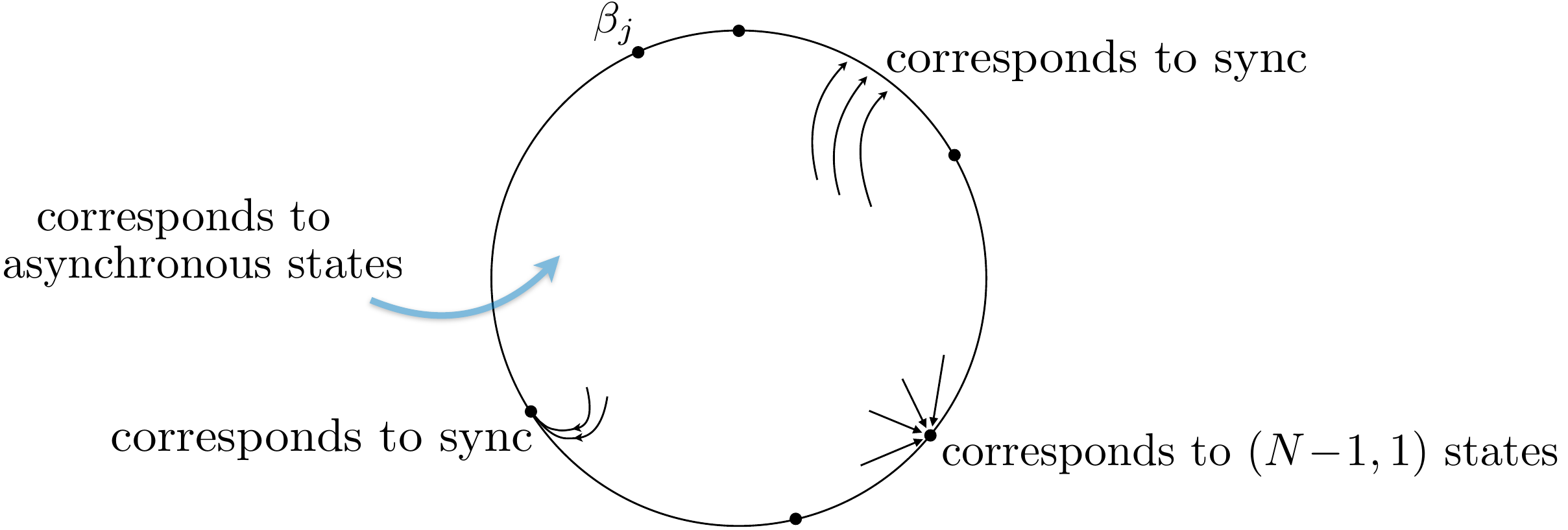}}
\caption{\label{fig1}
Points $w$ in $\Delta$ correspond to asynchronous states in \Tnmo.
As $w\to\beta\in\partial\Delta$,
$\beta\ne\beta_j$, the corresponding states in \Tnmo\ approach sync;
this also holds if
$w\to\beta_j$ tangentially.
As $w\to\beta_j$
at a fixed non-tangential angle $\phi$, 
the corresponding states in \Tnmo\ approach 
the \nnmo\ state with $\theta_j$ out of phase by $2\phi$.
}
\end{figure}

The boundary correspondence is illustrated in
Figure~\ref{fig1}.
A trajectory $w(t)$ approaching a point on the boundary of the disc distinct from any
$\beta_j$ 
(or approaching a $\beta_j$ tangent to the boundary circle)
corresponds to a trajectory approaching sync in \Tnmo.
A trajectory approaching a base point coordinate $\beta_j$
at a fixed angle $\phi$ corresponds a trajectory to approaching the \nnmo\ state with oscillator
$\theta_j$ phase shifted by $2\phi$ relative to the $N-1$ synchronized oscillators.
We have shown
previously~\cite{engelbrecht2014classification}
that any attracting or repelling states in \Tnmo\ 
of a system of the form (\ref{KsysC}) must be sync or \nnmo\ states,
which are the
the common boundary of asynchronous $G$-orbits.
So trajectories $w(t)$ approaching the boundary circle are particular particularly important to
understanding the dynamics of (\ref{KsysC}).

\section{$N = 2$ Dynamics}

The boundary dynamics for the system (\ref{AKS}) correspond to the dynamics for the system \ref{KsysC}) with $N = 2$, which is
$$
\begin{array}{rcl}
\dot z_1 & = & i \bigl( \omega + \Im (\A \overline z_1) \bigr )z_1\cr
\dot z_2 & = & i \bigl( \omega + \Im (\A \overline z_2) \bigr )z_2\cr
\end{array}
$$
with $z_1, z_2 \in S^1$ and $\A = s_1z_1 + s_2z_2$ for constants $s_1, s_2 \in \Bbb C$.
Since we are mainly interested in the phase between $z_1$ and $z_2$, we let
$\zeta = z_1 \overline z_2$ and analyze the dynamics of $\zeta$.  Note that if $\dot \zeta = 0$, then $\Im (\A\overline z_1)$ and $\Im(\A \overline z_2)$ depend only on the constant $\zeta$, and must be equal; otherwise we can see from the above equations that the phase between $z_1$ and $z_2$ would change in time.  So solutions to $\dot \zeta = 0$ correspond to solutions $(z_1, z_2)$ which are rotating at the same constant angular velocity.  The evolution equation for $\zeta$ is
$$
\begin{array}{rcl}
\dot \zeta &=& \dot z_1 \overline z_2 + z_1 \dot{\overline z}_2 \cr
&=& i\bigl( \omega + \Im(\A\overline z_1) \bigr) \zeta -  i \bigl( \omega + \Im (\A \overline z_2) \bigr ) \zeta \cr
&=& i \Im \bigl ( s_1 + s_2 \overline \zeta - s_1 \zeta - s_2 \bigr) \zeta \cr
&=& i \Im \bigl ( s_1 + \overline s_2  - s_1 \zeta -\overline s_2 \zeta  \bigr) \zeta \cr
 &=& i \Im \Bigl( ( s_1+ \overline s_2)(1 - \zeta)\Bigr) \zeta.
\end{array}
$$

We see that $\zeta = 1$, which corresponds to sync, is always a fixed point.  If we express $\zeta = e^{i \psi}$, then $\dot \psi = -i \overline \zeta \dot \zeta$, and we obtain the equivalent equation
\begin{equation}\label{BDyn}
\begin{array}{rcl}
\dot \psi &=& \Im \bigl( (s_1 + \overline s_2)(1-\zeta)\bigr) \cr
&=& -\Re(s_1+  s_2) \sin \psi + \Im (s_1 -  s_2) (1- \cos \psi). 
\end{array}
\end{equation}

This flow always has fixed point $\psi = 0$, corresponding to sync, with eigenvalue $-\Re (s_1+s_2)$.   If $\Re(s_1+  s_2) \ne 0$, then there is an additional fixed point $\psi^\ast \ne 0$ with opposite eigenvalue $\Re(s_1+s_2)$; $\psi = 0$ is stable if $\Re (s_1 +  s_2) > 0$, unstable if $\Re (s_1 +  s_2) <0$, and $\psi^\ast$ has the opposite stability.
  If $s_1 + \overline s_2 \ne 0$ but is pure imaginary, then $\psi = 0$ is the only fixed point, and is attracting globally but not locally near $\psi = 0$ (the flow has the same direction everywhere on the circle). The flow is identically $0$ iff $s_1 + \overline s_2 = 0$.  We mention in passing that the uniformly rotating solutions $(z_1, z_2)$ corresponding to fixed states $\zeta$ usually do not have angular velocity equal to $\omega$.  For example, the sync solution $z_1 = z_2$  has angular velocity $\omega + \Im(s_1 + s_2)$, which is not equal to $\omega$ unless $ \Im(s_1 + s_2)= 0$.

For any partition ${}\sim{}$ of $\{1, 2, \dots ,  N\}$ into two disjoint nonempty sets, the set of states $p = (\beta_1, \dots, \beta_N)$ where $\beta_j = \beta_k$  for $j \sim k$ is a 1D manifold in the reduced state space $T^{N-1}$,  invariant under the dynamics for any system (\ref{AKS}), and these 1D manifolds all meet at the sync state.  The dynamics on these two-cluster manifolds are given by the polar equation above, where $\psi$ measures the phase difference between the two clusters; the appropriate values of the coefficients $s_1, s_2$ are found by summing the $c_j$ over each of the two clusters.  The tangent directions to these 1D invariant manifolds at $\psi = 0$ are eigenvectors for the linearization of the system  at sync, and they span the full tangent space at sync.  Therefore we see from the polar equation above that the unique eigenvalue for the linearized dynamics at sync is $- \Re  (s_1 + \overline s_2) =- \Re c = -a$.  Consequently we see that {\sl the sync state in $T^{N-1}$  is linearly stable for  $a  > 0$, unstable for $a < 0$ and linearly neutral for $a = 0$}.  This is of course consistent with the much stronger result of Theorem 1, that sync is globally stable when $a>0$ and globally unstable when  $a < 0$.

\section{Fixed Point Analysis}

In this section we study the fixed points for the flow (\ref{weq}).  We begin with a lemma which will be crucial to the proofs of all our theorems.

\proclaim{Lemma 1.}{Assume that $N \ge 3$ and at least three $c_j \ne 0$.   Then the flow on the disc given by (\ref{weq}) has at most $(N-1)(N-2)$  fixed points.}

\pf The details are easier to carry out if we transform the $\dot w$ system to an equivalent system on the upper half plane, via the M\"obius transformations
$$
z = i{1-w \over 1+w}, \quad w = -{z-i \over z+i}
$$
which give a correspondence between the upper half plane $H = \{ \Im z > 0\}$ and the disc $\Delta = \{ |w| < 1\}$.  (Notice that $w = 0, 1, -1$ correspond to $z = i, 0, \infty$ respectively.)
Then the $\dot w$ flow transforms to
\begin{equation}\label{zeq}
\begin{array}{rcl}
\dot z ={dz \over dw} \dot w &=&\displaystyle -{2i \over (1+w)^2} \left (-{1 \over 2} (1- |w|^2)
 \sum_{j=1}^N c_j {\beta_j - w \over 1-\overline w \beta_j} \right)\cr
&=&\displaystyle{i \over 2} (z+i)^2 \left( -{1 \over 2}  \cdot  {|z+i|^2 - |z-i|^2 \over |z+i|^2}\right)
  \sum_{j=1}^Nc_j {\beta_j  + {z-i \over z+i} \over 1+  {\overline z+ i \over \overline z -i} \beta_j} \cr
&=&\displaystyle-{i \over 4} \left({ z+i \over \overline z -i}\right ) \cdot  4y \cdot
   \sum_{j=1}^N c_j{(z+i) \beta_j  + z-i \over  \overline z- i +(\overline z +i) \beta_j} \cdot {\overline z - i \over z+i} \cr
 &=&\displaystyle -i  y  \sum_{j=1}^Nc_j  {z-\alpha_j \over \overline z - \alpha_j}, \quad {\rm with} \ \ z = x+iy, \quad \alpha_j =  i{1-\beta_j \over 1+\beta_j} \in {\Bbb R}. 
\end{array}
\end{equation}
So we must prove that the equation
$$
\sum_{j=1}^Nc_j  {z-\alpha_j \over \overline z - \alpha_j} = 0
$$
has at most $(N-1)(N-2)$ solutions $z \in H$.  Let us assume without loss of generality that all $c_j \ne 0$, and $N \ge 3$.  Observe that
$$
\begin{array}{rcl}
\displaystyle
\sum_{j=1}^Nc_j  {z-\alpha_j \over \overline z - \alpha_j} = 0 
&\iff &\displaystyle
 \sum_{j=1}^Nc_j  {z-\overline z + \overline z - \alpha_j \over \overline z - \alpha_j} = 0 \cr
&\iff &\displaystyle
 (z - \overline z) \sum_{j=1}^N {c_j \over \overline z - \alpha_j}  + \sum^N_{j = 1}c_j = 0 \cr
&\iff &\displaystyle
  (z - \overline z) \sum_{j=1}^N {\overline c_j \over  z - \alpha_j} = \overline c.
\end{array}
$$
Let 
$$
F(z) = \sum_{j=1}^N {\overline c_j \over  z - \alpha_j},
$$
which is a rational function in $z$ with degree $N$.   The equation for fixed points  $z \in H$ is equivalent to $(z - \overline z) F(z) = \overline c$.  Express 
$$
F(z) = {P(z) \over Q(z)}, \quad {\rm where} \quad Q(z) = \prod_{j = 1}^N (z - \alpha_j)
$$
and $P(z)$ is a polynomial in $z$ with degree at most $N-1$. Since $F(z)$ has poles at $z = \alpha_j$, $P(z)$ and $Q(z)$ have no common factors.  Note that
$$
\lim_{z \to \infty} z F(z) = \overline c = \lim_{z \to \infty} {z P(z) \over Q(z)},
$$
which shows that the degree of $P(z)$ is exactly $N-1$, unless $c = 0$ and then the degree is at most $N - 2$.  Therefore we are done if $c = 0$; the equation $F(z) = 0$ can have at most $N - 2$ roots in $H$.

If $c \ne 0$, then WLOG we can set $c = 1$ and rewrite the fixed point equation  for $z \in H$ in the form $z - F(z)^{-1} = \overline z$.   Observe that
$$
z - F(z)^{-1} = {z P(z) - Q(z) \over P(z)};
$$
both $zP(z)$ and $Q(z)$ are degree $N$ monic polynomials, since $\lim_{z \to \infty} zP(z)/Q(z) = 1$.  Therefore $zP(z) - Q(z)$ has degree at most $N-1$.  We also see that $zP(z)-Q(z)$ and $P(z)$ have no common factors, since any common factor would also divide $Q(z)$.  Therefore $z- F(z)^{-1}$ is a rational function with degree $N-1$.  Let
$$
\Phi(z) = \overline{z - F(z)^{-1}},
$$
which  is a rational function in $\overline z$.  Our fixed point equation is equivalent to $\Phi(z) = z$ for $z \in H$.  Any fixed point for the map $\Phi(z)$ is also a fixed point for the second iterate $\Phi^2(z)$, which is a rational function in $z$ with degree $(N-1)^2$, and therefore has at most $(N-1)^2 + 1$ fixed points (a rational function of degree $r \ge 2$ has at most $r+1$ fixed points as a map of the extended complex plane $\hat{\Bbb  C}$).  The map $\Phi(z)$ has $N$ fixed points at $z = \alpha_j \in \Bbb R$, since $F(z)$ has  a pole at $z = \alpha_j$.  Therefore the number of fixed points $z \in H$ is at most $(N-1)^2 +1 - N = (N-1)(N-2)$.
\qed

We remark that the bound $(N-1)(N-2)$ is far from sharp; using the Lefschetz fixed point theorem and some other results from the theory of iterated 
rational maps~\cite{beardon2000iteration}, 
we can improve the bound to $4N-10$ (which coincidentally agrees with the previous bound for $N=3, 4$).  Since we only need the finiteness of the number of fixed points, we omit the proof, which would take us somewhat far afield.  

The condition that at least three $c_j \ne 0$ is necessary to insure finitely many fixed points; to see this, 
suppose only $c_1, c_2 \ne 0$ and all other $c_j=0$.  Then the fixed points for (\ref{weq}) in the disc are given by an equation of the form
$$
M_w(\beta_2) = \xi M_w(\beta_1)
$$
for some nonzero $\xi \in \Bbb C$.  Since $|M_w(\beta)| = 1$ for any $w \in \Delta$ and $\beta \in S^1$, we see that we must have $|\xi| = 1$ to have any solutions.  We also must have $\xi \ne 1$ to have solutions, since the map $M_w$ is one-to-one.  For $\xi \in S^1, \xi \ne 1$, the equation above is the equation of a circular arc in $\Delta$ joining the points $\beta_1$ and $\beta_2$.  When $\xi = -1$, this arc is the unique geodesic joining $\beta_1$ and $\beta_2$ for the hyperbolic metric on the disc, which we will discuss later.  For other $\xi$, these circular arcs form the family of curves of constant curvature joining $\beta_1$ and $\beta_2$, which are called hypercircles in hyperbolic geometry.  

Next, we study the linear stability of fixed points for the system (\ref{weq}).  The choice of coordinate $w$ depends on the base point $p \in T^N$, which without loss of generality we can choose so that $p$ is a fixed point for the flow in the reduced phase space $T^{N-1}$.   
With this choice
the flow (\ref{weq}) has fixed point at $w = 0$, and so $\sum c_j \beta_j = 0$.  To first order in $w$, the flow is given by
$$
\dot w = -{1 \over 2} \sum_{j = 1}^N c_j (\beta_j - w)(1 + \overline w \beta_j) = {1 \over 2} \Bigl( c w - Z_2 \overline w\Bigr), \quad {\rm where} \quad Z_2 = \sum_{j = 1} ^N  c_j \beta_j^2.
$$
If we write $w = u+iv$, $Z_2 = X_2 +iY_2$ and as usual $c = a + ib$,  then the 2D linear system for $\dot u, \dot v$ has matrix
$$
M = {1 \over 2} \pmatrix { a - X_2 & -b - Y_2 \cr b - Y_2 & a + X_2},
$$
which has ${\rm tr} \,  M = a$, $\det M = {1 \over 4} (|c|^2 - |Z_2|^2)$, and eigenvalues 
\begin{equation}\label{ev}
\lambda_\pm = {a \pm \sqrt{|Z_2|^2 - b^2} \over 2}. 
\end{equation}

Now suppose $a > 0$; then the fixed point $w = 0$ must have at least one eigenvalue $\lambda$ with $\Re \lambda >0$, and hence is a repelling node or spiral, a saddle, or a non-hyperbolic fixed point with one positive and one zero eigenvalue.  In the first two cases there are respectively $0$ or $2$ 
trajectories 
$w(t)$  that converge to the fixed point $w = 0$ as $t \to \infty$.  In the non-hyperbolic case with one $\lambda = 0$, if the fixed point at $0$ is isolated (which is the case if at least three $c_j \ne 0$) then there are at most two trajectories $w(t) \to 0$ as $t \to \infty$  
(see~\cite{perko2013differential}\ Section~2.11, Theorem~1).  
Therefore assuming at least three $c_j \ne 0$, we can conclude that {\sl in all cases there are at most two trajectories $w(t)$ that converge to the fixed point $0$ as $t \to \infty$.}  

When $a = 0$ the eigenvalues have the form $\pm \lambda$ with $\lambda$ either real or pure imaginary, so  fixed points can never be attracting in this case, and as we shall prove in the discussion preceding Theorem 2, can only attract finitely many trajectories. The case $a < 0$ is equivalent to the case $a > 0$ with time reversed.  The eigenvalues at the fixed point $w = 0$ are completely determined by the quantities $c$ and $Z_2$.  More generally, if $w  \in \Delta$ is a fixed point for the flow (\ref{weq}), the eigenvalues at $w$ are given by (\ref{ev}) with
$$
Z_2 = Z_2(w) = \sum_{j=1}^N c_j M_w (\beta_j)^2.  
$$

Now we consider the question of constraints on the number and type of fixed points for (\ref{weq}).
The equations for $w$ to be a fixed point for (\ref{weq}) with prescribed value $Z_2(w) = \xi$ are
$$
\sum_{j=1}^N c_j M_w (\beta_j) = 0, \quad \sum_{j=1}^N c_j M_w(\beta_j)^2 = \xi.
$$

Suppose we fix distinct $w_k \in \Delta$ and $\xi_k \in \Bbb C$ for $k = 1, \dots, r$, and also fix $c=a+ib$.  We wish to find coefficients $c_j$ such that $\sum c_j = c$ and (\ref{weq}) has fixed points at $w_k$ with $Z_2(w_k) = \xi_k$. This is a system of $2r+1$ linear equations in the $N$ coefficients $c_j$:
$$
\sum_{j= 1}^N c_j = c, \quad \sum_{j=1}^N c_j M_{w_k} (\beta_j) = 0, \quad \sum_{j=1}^N c_j M_{w_k} (\beta_j)^2 = \xi_k, \quad k = 1, \dots, r.
$$
We claim this system has solutions $c_j$ if $N \ge 2r+1$.  To prove this, consider the associated homogeneous system with $N=2r+1$.  If we transform to the upper half plane, as we did in the proof of Lemma 2, with $z_k = i{1-w_k \over 1+w_k} \in H, \alpha_j = i{1 - \beta_j \over 1+\beta_j} \in \Bbb R$, then the homogeneous system in the $c_j$ is equivalent to
$$
\sum_{j=1}^N c_j = 0, \quad \sum_{j=1}^Nc_j  {z_k-\alpha_j \over \overline z_k - \alpha_j} = 0,  \quad \sum_{j=1}^Nc_j \left( {z_k-\alpha_j \over \overline z_k - \alpha_j}\right)^2 = 0, \quad k = 1,  \dots, r.
$$
Using the identity
$$
{z-\alpha \over \overline z - \alpha} = 1 + {z-\overline z \over \overline z - \alpha}
$$
 together with $\sum c_j = 0$, we see that the homogeneous system is equivalent to the system
$$
\sum_{j=1}^N c_j = 0, \quad \sum_{j=1}^N{ c_j   \over \overline z_k - \alpha_j} = 0,  \quad \sum_{j=1}^N {c_j  \over (\overline z_k - \alpha_j)^2} = 0, \quad k = 1,  \dots, r.
$$
This implies that the rational  function $F(z)$ in Lemma 2 has $F(z_k) = F'(z_k) = 0$ for $k = 1, \dots, r$, which implies that its numerator $P(z)$ has $r$ double roots at the $z_k$.  If $F(z)$ is not identically $0$, then we must have $2r \le \deg P(z) \le N-2$, which is a contradiction.  Hence $F(z)$ is identically $0$, which means that all $c_j = 0$.  Therefore the inhomogeneous system has a unique solution for $N = 2r+1$, and infinitely many solutions for $N > 2r+1$.  In other words, we can find systems of the form (\ref{weq}) with as many fixed points as we desire, and can even prescribe the eigenvalues at the fixed points as we like, within the constraints imposed by the form of the eigenvalue equation (\ref{ev}).

\section{Boundary Flow Analysis}

To fully understand the dynamics of (\ref{weq}), we need to analyze the flow near the boundary of the disc $\Delta$.  
Consider first any point $\beta \in S^1$ which is distinct from any of the $\beta_j$.  Near $\beta$, the trajectories are the same curves as for the modified flow without the factor $(1/2)(1 - |w|^2)$, given by
$$
\dot w =- \sum_{j =1}^Nc_j  {\beta_j - w \over 1 - \overline w \beta_j}.
$$
This modified flow extends to a smooth flow on any open subset of $\Bbb C$ which excludes the $\beta_j$.  If $|w| = 1$, then
$$
\dot w = -w  \sum_{j =1}^N c_j {\overline w \beta_j - 1 \over 1 - \overline w \beta_j} =  c w .
$$
So we see that if $a = \Re c > 0$, then this flow points outwards and crosses the circle at every point $\beta \ne \beta_j$.  Hence the original $\dot w$ flow will have a unique trajectory converging in forward time to each $\beta \ne \beta_j$.  If $w$ is a point on this trajectory, then the forward limit set $\Omega_+(w)$ must be the single point $\{\beta\}$.  Similarly if $a < 0$, then there is a unique trajectory converging in backward time to each $\beta \ne \beta_j$.  We also see that if $a = 0$ but $b = \Im c \ne 0$, then no trajectory can converge in forward or backward time to any $\beta \ne \beta_j$, since the modified flow has trajectories along the arcs of the circle obtained by removing the $\beta_j$.

Next, we analyze the flow near the $\beta_j$.  As in the proof of Lemma 1, it is easier to transform the system to the upper half plane $H$ and study the $\dot z$ equation (\ref{zeq}).  Assume WLOG that $\beta_1 = 1$, so $\alpha_1 = 0$ and the remaining $\alpha_j \ne 0$.  We wish to analyze the flow near $z = 0$, and to do this we will employ the polar representation $z = r e^{i \theta}$, where $0 \le \theta \le \pi$.  We see that
\begin{equation}\label{zdot}
\begin{array}{rcl}
\dot z \overline z &=&\displaystyle  -i  y \left( c_1 z + \overline z\sum_{j=2}^N c_j {z-a_j \over \overline z - a_j},\right) \cr
&=&\displaystyle  -i  y \Bigl( c_1 z + (c-c_1) \overline z +O(r^2) \Bigr ) \cr
&=&\displaystyle  y\left (-i  c x + (2c_1-c) y + O(r^2), \right) 
\end{array}
\end{equation}
where $O(r^2)$ represents sums of terms of the form $r^k \cos m \theta, r^k \sin n \theta$ with $k \ge 2 $.
Using the polar conversions
$$
r \dot r = \Re(\dot z \overline z), \quad r^2 \dot  \theta = \Im( \dot z \overline z),
$$
we get the equivalent polar system
\begin{equation}\label{req}
\begin{array}{rcl}
\dot r  &=&  r  \sin \theta  \bigl ( b \cos \theta  +(2a_1-a)\sin \theta + O(r) \bigr ) \cr
\dot \theta &=& \sin \theta \bigl(  -a \cos \theta +(2b_1-b)  \sin \theta + O(r) \bigr).
\end{array}
\end{equation}
When $r = 0$,  the equation for $\dot \theta$ can be written in the form
$$
2 \dot \theta = -a  \sin 2 \theta +(2b_1-b) (1 - \cos 2 \theta),
$$
which is equivalent to (\ref{BDyn}) if we let $\psi = 2 \theta$ and $s_1 = c_1, s_2 = c- c_1$.  This makes sense, because the points with $r = 0$ correspond to the points on the \nnmo\ boundary component with all oscillators in sync except the first.  This equation has a unique fixed point $\theta^\ast \in (0, \pi)$, defined by
$$
\tan \theta^\ast= {a \over 2b_1-b},
$$
provided $a \ne 0$.  (There are no solutions in $(0,\pi)$  if $a = 0$, unless we also have $2b_1-b  = 0$; in this case, $\dot \theta = 0$ for {\sl all} $\theta \in (0,\pi)$.)  As we saw earlier, when $a \ne 0$, the fixed point $\theta = 0$ has eigenvalue $-a$ and the fixed point $\theta^\ast$ has eigenvalue $a$, in the direction along the interval with $r = 0$.   The linearization of the $\dot r$ equation at $r = 0, \theta = \theta^\ast$ is
\begin{equation}\label{eqrlin}
\begin{array}{rcl}
\dot r  &=&  r  \sin \theta^\ast  \Bigl ( b \cos \theta^\ast  +(2a_1-a)\sin \theta^\ast  \Bigr ) \cr
&=&\displaystyle
 r  \sin \theta^\ast  \Bigl ( b \left({2b_1-b \over a}\right) \sin\theta^\ast +(2a_1-a)\sin \theta^\ast  \Bigr ) \cr
&=&\displaystyle
{ r  \sin^2 \theta^\ast \over a}  \Bigl( b(2b_1-b) + a(2a_1-a) \Bigr) \cr
&=&\displaystyle
{ r  \sin^2 \theta^\ast \over a}  \Bigl(2 \Re (c\overline c_1) - |c|^2 \Bigr), \cr
\end{array}
\end{equation}
which is independent of $\theta$.

Now suppose $a > 0$.  Then the fixed point $r, \theta \ = 0, \theta^\ast$ has positive eigenvalue $a$, so must be a repelling node, a saddle, or a non-hyperbolic fixed point with exactly one non-zero eigenvalue. If we assume that at least three $c_j \ne 0$, then this fixed point is isolated.  If it is a repelling node then no trajectory $z(t)$ converges to the fixed point in forward time.  A saddle has two attracting trajectories, but they must be on opposite sides of the repelling trajectories (unstable manifolds) along the $\theta$ axis, so there is one attracting trajectory with $r > 0$. If the fixed point has a zero eigenvalue, it must be a repelling node, topological saddle or saddle-node 
(see~\cite{perko2013differential}\ Section~2.11, Theorem~1),
and in each case has at most one attracting trajectory with $r > 0$.  {\sl Therefore in all cases there is at most one trajectory converging to this fixed point in forward time.}   In backward time, a set of initial conditions with positive measure will converge to this fixed point if it is a repelling node; no trajectories converge to this fixed point if it is a saddle or topological saddle (the unstable manifolds have $r=0$), and at most a single trajectory with $r > 0$ can converge to the fixed point if it is a saddle-node.  These results will be crucial for the proof of Theorem 1.  

\section{Hyperbolic Geometry and the Gradient and Hamiltonian Conditions}

The factor $1-|w|^2$ in the $\dot w$ equation (\ref{weq}) suggests that this flow has connections to hyperbolic geometry.  The Poincare model for hyperbolic geometry on the unit disc $\Delta$  has metric
$$
ds = {2 | dw| \over 1-|w|^2}.
$$
This metric is conformal with the Euclidean metric (i.e.~angle measures agree), has constant negative curvature $-1$, and its geodesics are lines or arcs of circles which meet the boundary in $90^o$ angles. Since the reduced $G$-orbits are in one-to-one correspondence with $\Delta$ via the coordinate $w$, we can transfer this metric to the reduced $G$-orbits.  In fact, as shown 
in~\cite{chen2017hyperbolic},
this metric on the reduced $G$-orbits is {\sl independent} of the choice of base point.

In 2D Riemannian geometry the simplest flows are given by gradient and Hamiltonian vector fields.  A gradient vector field has the form $\nabla \Phi$ for some smooth real function $\Phi$, where the gradient is defined in terms of the Riemannian metric.  A Hamiltonian flow is the $90^o$ rotation of a gradient field $\nabla \Phi$, and therefore has $\Phi$ as a conserved quantity. The hyperbolic gradient of a function $\Phi (w)$ in complex form is given by
$$
\nabla_{hyp} \Phi (w) = \lambda^{-2} \nabla_{euc} \Phi (w) = 2 \lambda^{-2} {\partial \Phi \over \partial \overline w},
$$
where $\lambda = 2(1-|w|^2)^{-1}$ is the hyperbolic metric factor,  $\nabla_{euc}$ is the ordinary Euclidean gradient  and
$$
{\partial \over \partial \overline w} = {1 \over 2} \left ( {\partial \over \partial u} + i{\partial \over \partial v} \right), \quad w = u+iv.
$$
In~Ref~\cite{chen2017hyperbolic}
we derived criteria for the $\dot w$ flow to be gradient or Hamiltonian:  define the differential operator $D$ on the torus $T^N$ with coordinates $z_j \in S^1$ by
 $$
 D = {\partial \over \partial z_1} + \cdots + {\partial \over \partial z_n}.
 $$
 Then the $\dot w$ flow is gradient for the hyperbolic metric on all reduced $G$ orbits iff $\Im D \A = 0$ everywhere on $T^N$, and similarly is Hamiltonian iff $\Re D\A = 0$ everywhere on $T^N$.  For the asymmetric Kuramoto-Sakaguchi model with order parameter (\ref{AOP}), these conditions reduce to
 $$
\begin{array}{|rcl|}
\hline
 {\rm gradient} &\iff& b = 0 \cr
 {\rm Hamiltonian} &\iff&  a = 0
\\ \hline
\end{array}
 $$
In particular, we see that the symmetric K-S model, which has all $c_j = Ke^{i \psi}/N$, is gradient iff $\psi = 0$ or $\pi$, and Hamiltonian iff $\psi = \pm \pi/2$
(as first pointed out in Ref.~\cite{watanabe1994constants}).

We showed in~\cite{chen2017hyperbolic}
that the $\dot w$ flow for the symmetric K-S model with $K=1, \psi = 0$ is the hyperbolic gradient flow for the function
$$
\Phi (w) = -\log(1-|w|^2) + {2 \over N} \sum_{j = 1}^N \log |w-\beta_j|.
$$
It is not hard to modify this function to find the corresponding potential for the asymmetric case, assuming $c = a$ is real.  Observe first that
$$
\nabla_{hyp} \log (1 - |w|^2) = {1 \over 2} (1-|w|^2)^2 {\partial \over \partial \overline w} \log (1-w \overline w) = -{1 \over 2} w(1-|w|^2).
$$
We will need the identity
$$
{1 \over \overline w - \overline \beta} = -(1-|w|^2)^{-1} (w + M_w( \beta)),
$$
which follows from 
$$
M_w ( \beta) = {\beta -w \over 1-\overline w \beta} =  {1 - w (\overline \beta -\overline w) - |w|^2  \over \overline \beta  - \overline w} 
=  -w + {1  - |w|^2  \over \overline \beta - \overline w}.
$$
Using this, we see that
$$
\begin{array}{rcl}
\nabla_{hyp} \log |w - \beta | &=&\displaystyle
 {1 \over 4} (1-|w|^2)^2 {\partial \over \partial \overline w} \left ( \log (w-\beta)+ \log (\overline w - \overline \beta) \right)  \\[7pt]
&=&\displaystyle
  {1 \over 4} (1-|w|^2)^2 {1 \over \overline w -\overline \beta} 
= -{1 \over 4} (1-|w|^2) (w+M_w ( \beta)).
\end{array}
$$
Since $\log (w-\beta) = \log |w-\beta | + i \arg (w-\beta)$ is holomorphic,  $\nabla_{hyp} \log (w - \beta) = 0$, and therefore
$$
 \nabla_{hyp} \arg(w-\beta) = i \nabla_{hyp} \log |w - \beta | = -{i \over 4} (1-|w|^2) (w+M_w (\beta)).
$$

Using these ingredients, we can easily assemble a potential function $\Phi$ for the $\dot w$ flow when $c$ is real; using $c_j = a_j +i b_j$,
we 
see that
$$
\begin{array}{rcl}
\Phi(w) &=&\displaystyle -a \log(1-|w|^2) + 2 \sum_{j =1}^N a_j \log |w-\beta_j | + b_j \arg(w-\beta_j) \cr 
&=&\displaystyle -a \log(1-|w|^2) + 2 \Re \sum_{j =1}^N \overline c_j \log (w-\beta_j )
\end{array}
$$
satisfies
$$
\nabla_{hyp} \Phi(w) = -{1 \over 2} (1-|w|^2) \sum_{j = 1}^N c_j M_w( \beta_j),
$$
as desired.

Now suppose $c = a+ib$ is not real; then we can express $c = e^{i \psi} |c|$, and find a potential $\Phi$ for the $\dot w$ flow with coefficients $e^{-i \psi} c_j$.  Then the original $\dot w$ flow can be expressed as a ``twisted'' gradient flow:
$$
\dot w = e^{i \psi} \nabla_{hyp} \Phi(w).
$$
The quantity $\Phi(w)$ is now strictly increasing or decreasing along all non-trivial trajectories, unless $\psi = \pm \pi/2$; this is because
$$
\dot \Phi(w) = \langle \nabla_{hyp} \Phi(w) , \dot w \rangle_{hyp} = \langle \nabla_{hyp} \Phi(w) , e^{i \psi} \nabla_{hyp} \Phi(w) \rangle_{hyp} =\cos \psi  || \nabla_{hyp} \Phi(w) ||^2_{hyp}.
$$
When $\psi = \pm \pi/2$, the quantity $\Phi(w)$ is conserved, and the flow has Hamiltonian $\pm \Phi$.

\section{Dynamics Of The Asymmetric K-S Model: General Case}

The gradient / Hamiltonian structure described above makes it possible to give a fairly complete description of the dynamics of the asymmetric K-S model (\ref{AKS}) in the general case, which we state and prove in Theorem 1 in this section.  The key ingredient is the potential function $\Phi$ constructed in the previous section.  We will use the following lemma several times going forward.

\proclaim{Lemma 2.}{Suppose $\dot w = f(w)$ is a smooth flow on the disc $\Delta$ which has finitely many fixed points $w^\ast \in \Delta$, and there is a smooth function $\Phi$ on $\Delta$ such that $\dot \Phi > 0$ along all trajectories except fixed points.  Then for all $w \in \Delta$, the forward or backward limit set $\Omega_+(w), \Omega_-(w)$ is either a single fixed point $w^\ast \in \Delta$ or is completely contained in the boundary circle $S^1$.}
 
\pf  Let $w \in \Delta$ and assume $w$ is not a fixed point.  The forward limit set $\Omega_+(w)$ in the closed disc $\overline \Delta$ is nonempty, compact, connected and forward and backward invariant.  Suppose $\Omega_+(w)$ is not completely contained in $S^1$; let $w' \in \Omega_+(w) \cap \Delta$.   Then $\Phi (w(t) ) < \Phi(w')$ for all points $w(t)$ on the trajectory of $w$, and 
 $$
\Phi(w') = \lim_{t \to \infty} \Phi(w(t)).
 $$
If $w'$ is not a fixed point for the flow, then any forward time evolution $w''$ of $w'$ must also have 
 $$
 \Phi(w'') = \lim_{t \to \infty} \Phi(w(t)),
 $$
which is impossible since $\Phi(w') < \Phi(w'')$.  Hence $ \Omega_+(w) \cap \Delta$ must consist of finitely many fixed points in $\Delta$; since  $\Omega_+(w)$ is connected, this implies $\Omega_+(w) = \{w^\ast\}$ for a single fixed point $w^\ast$.
Clearly the same argument applies for the backward limit sets $\Omega_-(w)$.
\qed

Now we ready for the first main result in this paper.  
Note that the dynamics for the case $a = \Re c < 0$ are exactly the time reversal of the case for $a > 0$.

\proclaim{Theorem 1.}{Consider the system (\ref{AKS})  with $a = \Re c > 0$.  Then almost all trajectories in the reduced state space $T^{N-1}$ converge in forward time to sync and in backward time to a fully asynchronous state, or to one of finitely many  \nnmo\ states.}

\pf The result has been established above for $N=2$, so assume $N \ge 3$.  Choose any base point $p$ whose coordinates $\beta_j$ are all distinct.  Assume first that at least three of the $c_j$ are not $0$.  By Lemma 1, the associated $\dot w$ flow (\ref{weq})  has finitely many fixed points in $\Delta$.  Since $a > 0$, our fixed point analysis above shows that none of these can be attracting, and each can attract at most two trajectories.  Let $w \in \Delta$ and consider the forward limit set $\Omega_+(w)$.  If $w$ is on one of the finitely many trajectories converging to a fixed point $w^\ast$ in the disc, then $\Omega_+(w) = \{w^\ast \}$; otherwise by Lemma 2,   $\Omega_+(w)$ must be completely contained in the boundary $S^1$.  Suppose $\beta \in \Omega_+(w)$, and $\beta \ne \beta_j$.  Our analysis above of the dynamics near the boundary showed that then we must have $\Omega_+(w) =\{\beta\}$; this implies that the corresponding  trajectory in  $\widetilde{Gp}$ converges to sync.

The only other possibility in light of Lemma 2 is that $\Omega_+(w) = \{\beta_j\}$ for some $j$; in other words, the trajectory $w(t) \to \beta_j$ as $t \to \infty$. Let us convert this to a trajectory $z(t)$ in the upper half plane satisfying the equations (\ref{zdot}).   We claim that there is at most one trajectory $z(t)$ in the upper half plane that converges to 0;  this proves that the $\dot w$ flow has at most one  trajectory converging to each of the $\beta_j$.  Suppose $\lim_{t \to \infty} z(t) = 0$; then the corresponding polar coordinate $r(t) \to 0$ as $t \to \infty$.  Since $a > 0$, the $\dot \theta$ flow on the interval $(0,\pi)$ given by (\ref{req}) for $r = 0$ has a unique fixed point $\theta_j^\ast$.  Let $(\theta_0, \theta_1) \subset (0,  \pi)$ be any interval containing $\theta_j^\ast$.  Since $r(t) \to 0$, for $t$ sufficiently large the $\dot \theta$ flow on the intervals $(0, \theta_0]$ and $[\theta_1, \pi)$ converges to $0$ and $\pi$ respectively.    This means that $\theta(t)$ must converge to $0$ or $\pi$, or remain in $(\theta_0, \theta_1)$ as $t \to \infty$.  Since $(\theta_0, \theta_1)$ was arbitrary, in the latter case we have $\theta(t) \to \theta_j^\ast$.  Now if $\theta(t) \to 0$ or $\pi$, which are hyperbolic attracting fixed points for the 1D $\dot \theta$ flow, then $\sin \theta(t)$ is dominated by some decreasing exponential function $ K e^{-\mu t}$ for some $K, \mu > 0$ as $t \to \infty$.  Then the $\dot r$ equation in (\ref{req}) is dominated by  $K'r e^{-\mu t}$ for some other constant $K' > 0$.   Integrating the inequality $-\dot r \le K're^{-\mu t}$ from $0$ to $t$ gives
$$
-\log r(t) \le -\log r(0) + K'\left({1 - e^{-\mu t} \over \mu}\right),
$$
but this implies that $r(t)$ does {\sl not} decay to $0$ as $t \to \infty$.
Hence we must have $\theta(t) \to \theta_j^\ast$.  As we saw from the boundary flow analysis above, the fixed point $r, \theta = 0, \theta_j^\ast$ can attract at most one trajectory $(r(t), \theta(t))$.  This establishes the claim.

So we see that for any general base point $p$, all but finitely many trajectories on the reduced $G$-orbit $\widetilde{Gp}$ converge in forward time to the sync state, which proves the forward time assertion in the theorem.  Now let's consider backward limit sets $\Omega_-(w)$ for the $\dot w$ flow.  Clearly no trajectory $w(t)$ can converge to a boundary point $\beta \ne \beta_j$ as $t \to -\infty$, since the modified $\dot w$ flow, obtained by removing the scaling factor $1-|w|^2$, points outwards along the circle away from the $\beta_j$.  Therefore we must have $\Omega_-\{w\} = \{w^\ast\}$ for some fixed point $w^\ast \in \Delta$, or $\Omega_-\{w\} = \{\beta_j\}$ for some $j$.  In the first case, the corresponding trajectory on $\widetilde{Gp}$ is converging to a fully asynchronous state as $t \to -\infty$.  In the second case, the corresponding polar trajectory $r(t), \theta(t)$ has $r(t) \to 0$ as $t \to -\infty$.  An argument similar to the one above shows that $\theta(t) \to \theta_j^\ast$ as $t \to -\infty$:  we can rule out  $\theta(t) \to 0$ or $\pi$ as $t \to -\infty$, because then the corresponding trajectory on $\widetilde{Gp}$ converges to sync in backward time; this can't happen because sync is attracting.  This shows that all trajectories in $\widetilde{Gp}$ converge in backward time to a fully asynchronous state or to one of the \nnmo\ states corresponding to the $\theta^\ast_j$.

Finally, we consider the case where at most two of the $c_j$, say $c_1$ and $c_2$, are not $0$.  All the arguments above go through, unless $|c_1|^2 = |c_2|^2$ and the flow (\ref{weq}) has a circular arc of fixed points joining $\beta_1$ and $\beta_2$.  Assume $\beta_1 = 1, \beta_2 = -1$ and convert the flow to an equivalent flow on the upper half plane, as we did above in the boundary flow analysis.  Then $\alpha_1 = 0, \alpha_2 = \infty$ and the flow for $z \in H$ is given by $\dot z \overline z = -iy(c_1 z +  c_2 )$.  The equivalent polar system is
\begin{equation}\label{Req}
\begin{array}{rcl}
\dot r  &=&  r  \sin \theta  \bigl ( b \cos \theta  +(a_1-a_2)\sin \theta \bigr ) \cr
\dot \theta &=& \sin \theta \bigl(  -a \cos \theta +(b_1-b_2)  \sin \theta \bigr).
\end{array}
\end{equation}
The $\dot \theta$ equation has no $r$ dependence, and  has a unique repelling fixed point $\theta^\ast \in (0, \pi)$.  The relation $|c_1|^2 = |c_2|^2$ is equivalent to the relation
$$
a(a_1-a_2) =-b(b_1-b_2),
$$
which implies that $a\dot r / r = -b \dot \theta$, and so the flow has fixed points along the ray $\theta =\theta^\ast$.    Integrating this relation gives
$$
r(t) = r(0) \exp\bigl (-(b/a)(\theta(t) -\theta(0)\bigr). 
$$
Since $a > 0$, $\theta(t) \to 0$ or $\pi$ as $t \to \infty$ unless $\theta(0) = \theta^\ast$, and $\theta(t) \to \theta^\ast$ as $t \to -\infty$.  From this we see that $r(t)$ converges to some positive number as $t \to \pm \infty$.  All trajectories $z(t)$ off the ray of fixed points converge in forward time to some nonzero point on the real axis, and in backward time to some point on the ray of fixed points.  Hence almost all trajectories in $\widetilde{Gp}$ converge to sync in forward time and to some asynchronous state in backward time.  This completes the proof.
 \qed
In~\cite{lohe2017ws},
Lohe proves (in our notation) that 
when $a\ne0$, $|w(t)|$ converges; from this it is deduced that
in fact $|w(t)|$ converges.
This ignores the possibility that the limit set of the trajectory could be a circle of fixed points
in the interior of the disc, or the boundary circle.
Our Lemma~1 precludes the first possibility, and our analysis of the flow near the boundary precludes the 
second.
In addition, 
Lohe asserts that
if $|w(t)|\to1$ the corresponding trajectory in \Tnmo\ goes to sync.
This 
does not hold
if $w(t)\to\beta_j$ non-tangentially,
so one must rule out the possibility that a 
set positive measure of trajectories converge to some $\beta_j$,
which we did in our analysis of the flow at the boundary.

As we saw earlier, the \nnmo\ states corresponding to the angles $\theta_j^\ast$ do not depend on the choice of the base point; nor do the eigenvalues governing their stability.  If one of these \nnmo\ fixed points is repelling, which happens if $2 \Re(c \overline c_j) > |c|^2$, then this fixed point has a $t \to -\infty$ basin  with positive measure in the reduced state space $T^{N-1}$.  The asynchronous fixed points corresponding to fixed points $w^\ast \in \Delta$ {\sl do} depend on the base point $p$; since these are in the interior of the reduced $G$-orbits, they will form codimension two families of fixed points, which are neutrally stable in $N-3$ directions.  

\begin{figure}[b]
\centerline{\includegraphics[width=4.7in]{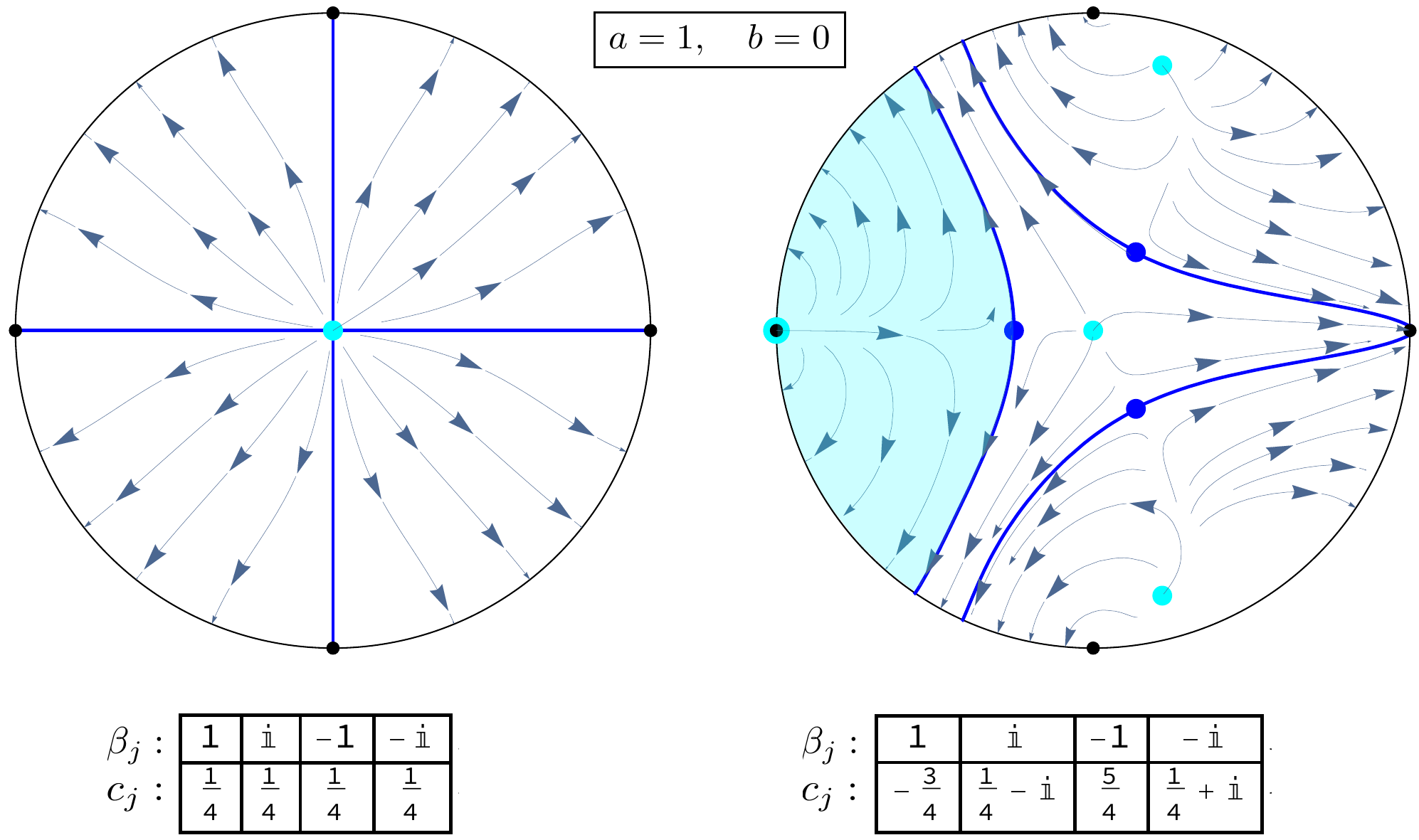}}
\caption{\label{fig2}
Two examples of $\dot w$ flows for $N=4$ and $c=1$ (stable sync); 
first panel is for the classic model with equal $c_j$.
Second panel is an example with
the maximum number 6 fixed points in the disc;
together with $w^*=-1$ there are 4 repellors (cyan dots) and
3 saddles (blue dots). Base point components are black dots and
the tables give values for $\beta_j$ and $c_j$.
}
\end{figure}

In
Figure~\ref{fig2} 
we contrast the $\dot w$ field for the symmetric Kuramoto model with
the $\dot w$ field for
an asymmetric model 
exhibiting more complicated dynamics 
consistent with Theorem 1.
We choose
$a=1$ (so sync is stable), $N=4$ 
and
base point $p=(1,i,-1,-i)$ 
for simplicity.
The first panel is for the classic model with all $c_j=\frac14$,
which has 
a repelling fixed point at $w=0$.
The second panel is for an example with
unequal $c_j$ chosen to give
the maximum number 6 of fixed points in the disc: 
3 are saddles (blue dots) and 
4 are repellors (cyan dots).
In both examples almost all trajectories converge in forward time to some point
on the boundary circle distinct from the $\beta_j$;
the corresponding trajectories in \Tnmo\ converge to sync.
In backward time,
trajectories with white background converge to a repellor in the disc,
which corresponds to an asynchronous state in \Tnmo.
In the second example,
trajectories with cyan background converge to 
the boundary point $\beta_3=-1$ 
(cyan dot) 
at angle $\pi/2$;
the corresponding trajectories in $T^{N-1}$ converge to
the \nnmo\ fixed point with $\theta_3$
out of phase by angle $\pi$.
In the first example 
the blue lines are exceptional;
they converge in forward time to
a boundary point $\beta_j$
at angle $\pi/2$;
the corresponding trajectories in $T^{N-1}$ converge to
the \nnmo\ fixed point with $\theta_j$
out of phase by angle $\pi$.
In the second example 
the blue lines are the separatrices between the
different $t\to-\infty$ basins of attraction.

\section{Dynamics Of The Asymmetric K-S Model: Hamiltonian Case}

Next we consider the dynamics in the case $a = 0$, when the $\dot w$ system (\ref{weq}) has Hamiltonian structure.  We begin with a lemma that is analogous to Lemma 2 in the previous section.

\proclaim{Lemma 3.}{Suppose $\dot w = i \nabla_{hyp} \Phi (w)$ is a Hamiltonian flow on the disc $\Delta$ which has finitely many fixed points $w^\ast \in \Delta$.  Then for all $w \in \Delta$, either its trajectory w(t) is a closed orbit, or $\Omega_+(w)$ is  a single fixed point $w^\ast \in \Delta$, or $\Omega_+(w)$ is completely contained in the boundary circle $S^1$.}
 
\pf  Let $w \in \Delta$, and suppose $w' \in \Omega_+(w) \cap \Delta$.  Since $\Phi$ is conserved for the flow, $\Phi(w') = \Phi(w)$.  If $w'$ is not a fixed point for the flow, then we can find a neighborhood $U \subset \Delta$ of $w'$ such that for $w'' \in U$, $\Phi(w'') = \Phi(w')$ if and only if $w''$ lies on the trajectory of $w'$.  The trajectory $w(t)$ of $w$ must enter $U$, and 
therefore we must have $w$ and $w'$ on the same trajectory: $w' = w(t_1)$ for some time $t_1$.  But $w' \in \Omega_+(w)$, which implies that we can find a sequence of times $t_n \to \infty$ with $w' = w(t_n)$; this implies that $w(t)$ is periodic.  The other possibilities are that $w'$ is a fixed point, or $\Omega_+(w) \subset S^1$.  In the first case, since there are finitely many fixed points and the limit set $\Omega_+(w)$ is connected, we must have $\Omega_+(w) = \{w^\ast\}$ for a single fixed point $w^\ast \in \Delta$.
\qed

Before we proceed with the proof of Theorem 2, we need a preliminary result about fixed points for 2D Hamiltonian flows.  Suppose $\dot w = i \nabla_{hyp} \Phi (w)$ is a Hamiltonian flow on the disc $\Delta$ which has finitely many fixed points $w^\ast \in \Delta$, and let us also assume that the Hamiltonian function $\Phi$ is real analytic.  Then we claim for any fixed point $w^\ast \in \Delta$ there are  at most finitely many trajectories $w(t) \to w^\ast$ as $t \to \infty$.  The eigenvalues at $w^\ast$ are of the form $\pm \lambda$, where $\lambda$ is real or pure imaginary.  If $\lambda > 0$ then $w^\ast$ is a saddle, which has exactly two attracting trajectories.  If $\lambda = \pm i \nu \ne 0$, then $w^\ast$ is a (nonlinear) center; this result needs the real analyticity of $\Phi$ 
(see~\cite{perko2013differential}\ Section~2.14, Theorem~2).
No trajectories can  converge to a center fixed point.  The following argument covers the degenerate case with $\lambda = 0$ a double eigenvalue (as well as the saddle and center cases). Choose $r > 0$ small enough so that the closed disc  $\overline \Delta _r(w^\ast)$ of radius $r$ around $w^\ast$ is contained in the unit disc $\Delta$ and contains no fixed points except $w^\ast$.  Suppose there are infinitely many distinct trajectories $w(t)$ converging to $w^\ast$ as $t \to \infty$.  Only finitely many of these trajectories can  intersect  $C_r =\partial \overline\Delta_r(w^\ast)$; otherwise the real analytic function $\Phi$ would take the same value $\Phi(w^\ast)$ at infinitely many distinct points on $C_r$, which implies $\Phi$ is constant on $C_r$.  But if $\Phi$ is constant on $C_r$, then $C_r$ must be a closed orbit, and then none of the trajectories converging to $w^\ast$ can intersect $C_r$.  Therefore there must be a trajectory $w(t)$ converging to $w^\ast$ with $w(t) \in \overline \Delta _r(w^\ast)$ for all $t$.  Lemma 3 implies that $w(t)  \to w^\ast$ also  as $t \to -\infty$; in other words, $w(t)$ is a homoclinic orbit to and from $w^\ast$.  This orbit together with $w^\ast$ forms a simple closed curve $\Gamma$ contained in $\overline \Delta _r(w^\ast)$.  Then $\Phi$ is constant on $\Gamma$ but cannot be constant on its interior; hence $\Phi$ has a critical point inside $\Gamma$, contradicting the assumption that there are no fixed points in $\overline \Delta _r(w^\ast)$ except $w^\ast$.

\proclaim{Theorem 2.}{Consider the system (\ref{AKS}) with $c = ib \ne 0$.  If $b_j \ne b/2$ for all $j$, then almost all trajectories in the reduced state space $T^{N-1}$ are periodic or homoclinic connections to and from sync. If some $b_j = b/2$, there is also a positive measure set of initial conditions with trajectories that converge in forward time to \nnmo\ states, and similarly a positive measure set of initial conditions with trajectories that converge in backward time to \nnmo\ states.}

\pf  The result has been established above for $N=2$, so assume $N \ge 3$.  We begin as in the proof of Theorem 1: choose any base point $p$ whose coordinates $\beta_j$ are all distinct.  Assume first that at least three of the $c_j$ are not $0$. By Lemma 1, the associated $\dot w$ flow (\ref{weq}) has finitely many fixed points in $\Delta$. The $\dot w$ flow (\ref{weq}) has Hamiltonian
$$
\Phi(w) = -b \log(1-|w|^2) + 2 \Re \sum_{j =1}^N i \overline c_j \log (w-\beta_j),
$$
 and we see that $\Phi(w) \to \pm \infty$ as $w \to \beta \in S^1$, $\beta \ne \beta_j$.  Since $\Phi$ is conserved along trajectories, 
Lemma 3 implies that all trajectories $w(t)$ are either periodic, or converge to fixed points $w^\ast \in \Delta$, or converge to one of the $\beta_j$.  From our discussion of fixed points above, we know that only finitely many trajectories can converge to  fixed points $w^\ast \in \Delta$.

It remains to analyze the behavior of trajectories with $w(t) \to \beta_j$.  As in the proof of Theorem 1, we assume $\beta_1 = 1$, and convert the system to the upper half plane $H$, via the M\"obius transformation $w = -{z-i \over z+i}$.  Then a trajectory $w(t) \to \beta_1 = 1$ transforms to  a trajectory $z(t) \to 0$ for the system (\ref{zdot}).  Observe that
$$
\begin{array}{rcl}
\log(1 - |w|^2) &=&\displaystyle
\log \left( 1 - \left| {z-i \over z+i} \right |^2 \right) = \log ( |z+i|^2 - |z-i|^2) - \log |z+i|^2 \cr
&=&\displaystyle
\log 4y - \log |z+i|^2 = \log y + f(z),
\end{array}
$$
with $z = x+iy$, $y > 0$  and $f(z)$ real analytic near $z = 0$.  We also have
$$
\begin{array}{rcl}
\log(w-1) &=&\displaystyle
 \log \left(  - {z-i \over z+i}  - 1\right)  = \log(-z+i-z-i) -\log (z+i)  \cr
&=& \log z + g(z), \cr
\end{array}
$$
with $g(z)$ holomorphic near $z = 0$. The corresponding Hamiltonian function $\Psi(z)$ has the form
$$
\begin{array}{rcl}
 \Psi(z) = \Phi(w) &=& -b \log y  +2 \Re ( i \overline c_1 \log z ) + h(z) \cr
 &=&-b \log y + 2 b_1 \log |z| -2a_1 \arg z +h(z) \cr
 &=&\displaystyle
 -b\left( \left( 1 - {2b_1 \over b} \right) \log r + \log \sin \theta \right ) -2a_1 \theta + h(z),
\end{array}
$$
where $z = re^{i \theta}$,   and $h(z)$ real analytic near $z = 0$.

Now $\log r \to -\infty$ as $r \to 0$, and $\log \sin \theta \le 0$ for all $\theta$.  So  if $b_1/b < 1/2$, then the coefficient of $\log r$ above is positive, and $\Psi(z) \to \pm \infty$ as $z \to 0$; hence no trajectory $z(t)$ can converge to $0$.  If $b_1/b > 1/2$, then $z(t) \to 0$ implies $\theta(t) \to 0$ or $\pi$; otherwise the Hamiltonian $\Psi(z)$ will diverge.  Therefore the trajectory $z(t)$ converges to $0$ tangent to the real axis, and the corresponding trajectory converges to sync in the reduced group orbit $\widetilde{Gp}$.  
So we see that if $b_j \ne b/2$ for all $j$, then  all but finitely many trajectories in the disc $\Delta$ are either periodic orbits, or converge in forward and backward time to (perhaps different) $\beta_j$;  the corresponding trajectories in $\widetilde{Gp}$ are periodic orbits or homoclinic connections to and from sync.  This proves the theorem in the case $b_j \ne b/2$.

Now suppose, say,  $b_1/b = 1/2$; then the Hamiltonian has the form
$$
\Psi(z) = \Psi(re^{i \theta}) = -b\left ( \log \sin \theta + {2a_1 \over b} \theta \right) + h(re^{i \theta})
$$
as $r = |z| \to 0$.  In polar coordinates $r, \theta$, $h(re^{i \theta})$ is constant on the interval $r = 0$, and the  function
$$
f (\theta) = \log \sin \theta + {2a_1 \over b} \theta
$$
has a unique critical point $\theta_0 \in (0, \pi)$,  which is the unique root of $b \cos \theta + 2a_1 \sin \theta =0$ in $(0, \pi)$.  Therefore the gradient of $\Psi$ in polar coordinates is nonzero at any point $r=0, \theta \ne \theta_0$, $\theta \in (0,\pi)$.  This implies that there is a smooth level set of $\Psi$ which meets $r=0$ at angle $\theta$.  This level set must contain a trajectory converging to $\theta$ in forward or backward time.  If we examine the $\dot r$ equation in (\ref{req}), we see that as $r \to 0$, $\dot r$ changes sign at the critical point $\theta_0$ of $u(\theta)$.  This means that the trajectories converge to angle $\theta$ in forward time on one side of $\theta_0$ and in backward time on the other side.  Therefore in the $w$ disc $\Delta$, we have a one-parameter family of trajectories converging to $\beta_1$ at all possible angles except perhaps $\theta_0$; the convergence is in forward time on one side of $\theta_0$ and in backward time on the other side. As we vary the base point $p$, these trajectories will form sets of positive measure in the reduced state space $T^{N-1}$, converging to \nnmo\ states in either forward or backward time.

Finally, we consider the case where at most two of the $c_j$, say $c_1$ and $c_2$, are not $0$.  As in the proof of Theorem 1, all the arguments above go through unless $|c_1|^2 = |c_2|^2$.  In this case the relations  $a(a_1-a_2) =-b(b_1-b_2)$, $a = 0$ and $b \ne 0$ imply $b_1 = b_2 = b/2$.  The equivalent flow on the upper half plane in  polar coordinates (\ref{Req}) has $\dot \theta = 0$ and hence $\dot r / r $ is constant.  All
trajectories are rays converging in forward or backward time to $0$ or $\infty$, except for the ray of fixed points $\theta = \theta^\ast$, where $\theta^\ast$ is the unique root of the $\dot r$ equation in (\ref{Req}).  The corresponding trajectories in $\widetilde{Gp}$ converge to \nnmo\ states. This completes the proof.
\qed

\begin{figure}[t]
\centerline{\includegraphics[width=6.7in]{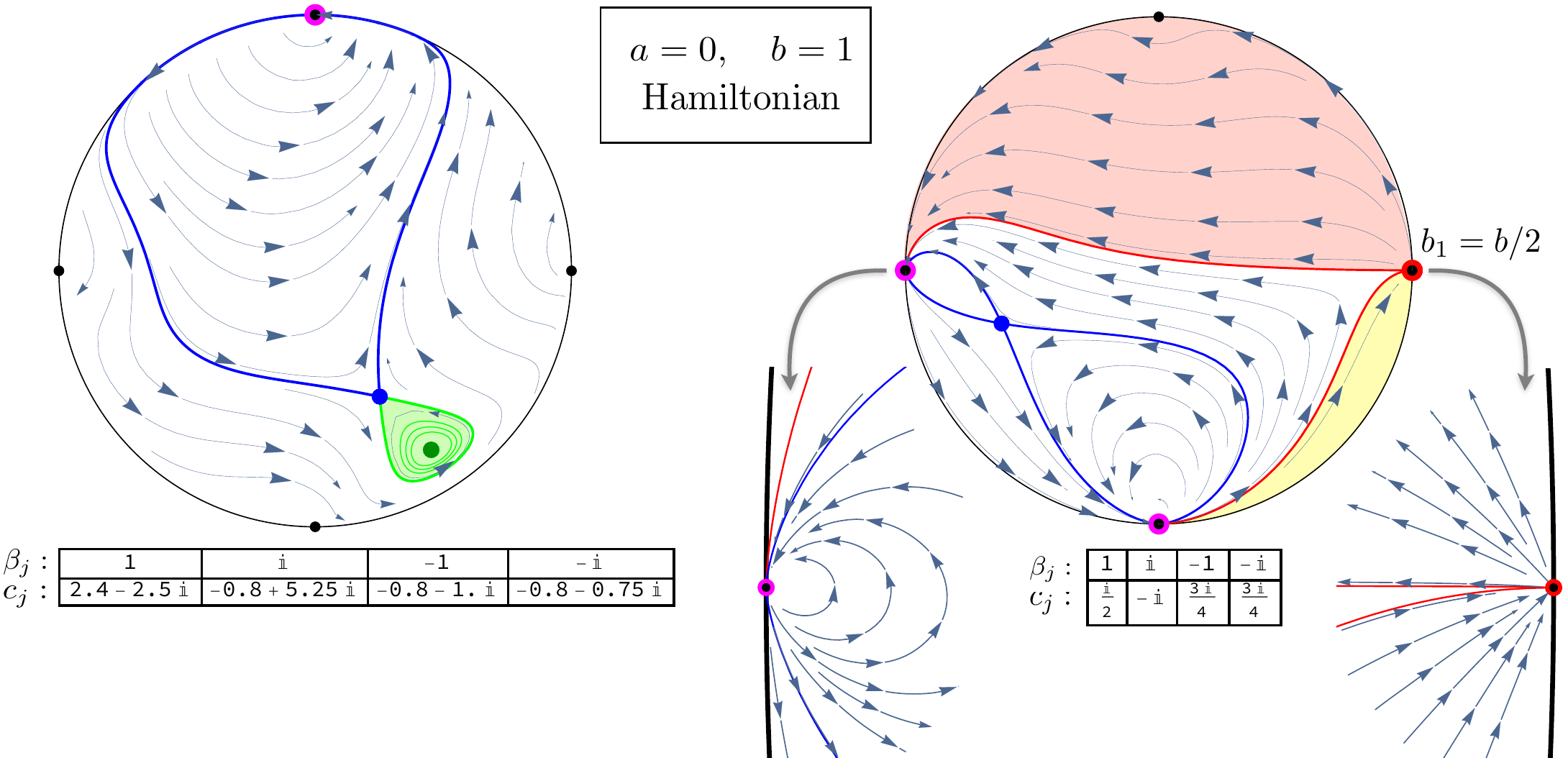}}
\caption{\label{fig3}
Two examples of Hamiltonian $\dot w$ flows for $N=4$. 
The trajectories with white background flow tangentially to the boundary
(corresponding to sync).
First panel has a center and a saddle, and no $b_j=b/2$.
Trajectories with a green background are periodic.
Second panel has weights $c_j$
chosen to satisfy $b_1=b/2$, and now
trajectories with 
yellow (red) backgrounds have 
forward (backward) \nnmo\ limit sets.
The two insets show the flow magnified near corresponding $\beta_j$.
}
\end{figure}

Figure~\ref{fig3} shows two $\dot w$ flows in the Hamiltonian case 
for $N=4$.
The trajectories with white and green backgrounds 
illustrate the generic dynamics in the first part of Theorem~2~($b_j\ne b/2$)
and the 
trajectories with 
yellow and red backgrounds illustrate the possibility of \nnmo\ forward or backward limits sets
when some $b_j=b/2$.
The first example has
a center (green) and a saddle (blue),
and
no $b_j=b/2$.
The periodic orbits (green region) are bounded by the homoclinic saddle connection (green).
Trajectories with white background 
approach $\beta_2=i$ tangentially, in forward and backward time;
the corresponding trajectories in \Tnmo\ are homoclinic connections
to/from sync.
Trajectories with green background are periodic orbits.
The blue trajectories are exceptional;
they correspond to
heteroclinic connections between the asynchronous saddle and sync in \Tnmo.
The second example has a saddle (blue)
and
$b_1=b/2$.
The two insets show the flow near $\beta_3$ and $\beta_1$. 
Trajectories with white background 
approach $\beta_3=-1$ or $\beta_4=-i$ tangentially, in forward and backward time;
the corresponding trajectories in \Tnmo\ are homoclinic connections
to/from sync.
Trajectories with red background 
approach $\beta_3=-1$ tangentially in backward time,
and
approach $\beta_1=1$ non-tangentially in forward time;
the corresponding trajectories in \Tnmo\ are heteroclinic connections
from sync to \nnmo\ states with $\theta_1$ out of phase.
Similarly trajectories with yellow background 
correspond to heteroclinic connections
from \nnmo\ states with $\theta_1$ out of phase to sync.
As in the first example,
the exceptional blue trajectories 
correspond to
heteroclinic connections between the asynchronous saddle and sync in \Tnmo.

\section{Special Case $\bf c=0$:  Gradient + Hamiltonian Dynamics}

In this section we consider the special case $c = 0$, which implies that the $\dot w$ flow (\ref{weq}) is simultaneously gradient and Hamiltonian with respect to the hyperbolic metric on $\Delta$.  What sort of flows have this dual gradient + Hamiltonian structure?  Such a flow has the form
$$
\dot w = 2 \lambda^{-2} {\partial U \over \partial \overline w} = -2i \lambda^{-2} {\partial V \over \partial \overline w},
$$
where $U$ and $V$ are smooth real functions on $\Delta$, and $\lambda$ is the hyperbolic metric factor.  The functions $U, V$ satisfy this relation if and only if 
$$
{\partial \over \partial \overline w} \left( U + i V \right) = 0,
$$
which is equivalent to the condition that the complex function $F = U + i V$ is holomorphic on $\Delta$.  In this case, with $w = u+iv$, 
$$
\dot w = \lambda^{-2} \left( {\partial U \over \partial u} + i {\partial U \over \partial v}  \right) = \lambda^{-2} \left( {\partial U \over \partial u} - i {\partial V \over \partial u}  \right) = \lambda^{-2} \overline{ F'(w)}
$$
from the Cauchy-Riemann equations, and we see that the fixed points of the flow correspond to zeroes of $F'(w)$.  
We call $F$ the {\sl holomorphic potential} for the flow.  
The holomorphic function $F$ on the disc is uniquely determined up to a constant by its real part $U$; in the case of the $\dot w$ flow  (\ref{weq}) for the K-S model with $c = 0$, we have
$$
U(w) = 2 \Re \sum _{j = 1}^N \overline c_j \log(w-\beta_j) \Longrightarrow F(w) = 2 \sum _{j = 1}^N\overline c_j  \log(w-\beta_j),
$$
where we take any single-valued branch of $\log (w-\beta_j)$ on the disc. With these preliminaries, it is fairly straightforward to describe the dynamics of the $\dot w$ flow.  We first consider the case when all $b_j \ne 0$.

\proclaim{Theorem 3.}{Consider the  K-S system with $c=0$ and  $b_j \ne 0$ for all $j$.    Then almost all trajectories in the reduced state space $T^{N-1}$ are homoclinic connections to and from sync.}

\pf  Assume $N \ge 3$ (the case $N=2$ follows directly from (\ref{BDyn}). The fixed points for (\ref{weq}) correspond to the roots of the equation $F'(w) = 0$ in $\Delta$, which we saw earlier has at most $N-2$ solutions in the disc.   If $w^\ast$ is a fixed point, then we can expand $F'(w)$ near $w = w^\ast$ in a power series
$$
F'(w) = a_k (w-w^\ast)^k + a_{k+1} (w-w^\ast)^{k+1}+ \cdots
$$
where $a_k \ne 0$ is the leading coefficient.  Therefore the leading term in the expansion of the flow for $w  = w^\ast + \eta$ is
$$
\dot \eta = \overline a_k \overline \eta^k,
$$
which is a (possibly higher order) saddle with index $-k$.   As such there will be $2k+2$ saddle trajectories converging to $w^\ast$ in forward or backward time, in an alternating arrangement around the saddle point.

The flow is gradient, so Lemma 1 implies that the limit sets $\Omega_+(w)$ and $\Omega_-(w)$ must be either a single fixed point $\{w^\ast\}$ in $\Delta$, or be completely contained in the boundary $S^1$.   The quantity $V(w) = \Im F(w)$ is conserved for the flow, so all trajectories lie on contours $\{V(w) = V(w_0)\}$.  We see that
\begin{equation}\label{Veq}
V(w) = 2\sum_{j = 1}^N \Bigl( a_j \arg (w - \beta_j) - b_j \log |w - \beta_j| \Bigr) , 
\end{equation}
which diverges as $w \to \beta_j$.  Hence no trajectories can converge to any $\beta_j$ if $b_j \ne 0$.    The limit set $\Omega_+(w)$ cannot be an arc on the circle, because the real-analytic function $V$ cannot be constant along any arc on the circle. So we see that all but finitely many trajectories in the disc $\Delta$  converge in forward and backward time to some point on the circle $\beta \ne \beta_j$;  the corresponding trajectories in $\widetilde{Gp}$ are  homoclinic connections to and from sync. 
\qed

There is no limit to the number of saddles for these flows; to see this, let $w_1, \dots, w_r$ be any points in $\Delta$.  The homogeneous linear system in $c_j$, $j = 1, \dots, N$, given by
$$
\sum_{j = 1}^N{c_j \over    \overline w_k -\overline \beta_j } = 0 \quad  k = 1,  \dots, r \quad {\rm and} \quad \sum_{j = 1}^N c_j = 0
$$
will have nontrivial solutions for $N \ge r+2$, so we can construct a flow of this type with fixed points at all $w_k$.  For a fixed set of coefficients $c_j$, the number of saddles can also vary as the coordinates of the base point vary.  For example, consider the $c_j$ given by $1, 1, -1, -1$.  For $\beta_j = 1, -1, i, -i$ the holomorphic potential $F(w)$ satisfies
$$
F'(w) = {2 \over w-1} + {2 \over w+1} - {2 \over w-i} - {2 \over w+i} = {8w \over w^4 - 1},
$$
which has a single zero at $w = 0$ in $\Delta$.  But if we switch to  $\beta_j = 1, i, -1, -i$, then
$$
F'(w) = {2 \over w-1} + {2 \over w-i} - {2 \over w+1} - {2 \over w+i} = {4(1+ i)(w^2 -i) \over w^4 - 1},
$$
which has no zeros in $\Delta$.  

If $b_j = 0$ for some $j$, then the dynamics can be more complicated; we will need to analyze the conserved quantity $V(w)$ more carefully.  
Observe that $F(w)$ is holomorphic along the boundary circle except at the $\beta_j$.  If $\beta \in S^1$, $\beta \ne \beta_j$, then the directional derivative of $F$ at $\beta$  in the direction $i \beta$ tangent to the circle at $\beta$ is
$$
(D_{i \beta} F) (\beta) = i \beta F'(\beta) = 2i \beta  \sum_{j = 1}^N {\overline c_j   \over \beta - \beta_j} =2 i\sum_{j = 1}^N  {\overline c_j  \beta_j \over \beta - \beta_j},
$$
because
$$ \sum_{j = 1}^N {\overline c_j  \beta \over \beta - \beta_j}  - \sum_{j = 1}^N  {\overline c_j  \beta_j \over \beta - \beta_j} =  \sum_{j = 1}^N \overline c_j { \beta - \beta_j \over \beta - \beta_j} =0.
$$
Observe that
$$
\overline{(D_{i \beta} F) (\beta)} = -2i \sum_{j = 1}^N { c_j \overline \beta \over \overline \beta - \overline \beta_j} \cdot {\beta_j \beta \over \beta_j \beta} =2 i \sum_{j = 1}^N {c_j \beta_j\over \beta - \beta_j}.
$$
Now $V(\beta) = \Im F(\beta)$ is constant for $\beta$ along some arc on the circle if and only if $(D_{i \beta} F) (\beta) \in \Bbb R$ along this arc, which is equivalent to
$$
\sum_{j = 1}^N {c_j \beta_j\over \beta - \beta_j}
 = \sum_{j = 1}^N {\overline c_j \beta_j\over \beta - \beta_j}
$$
along this arc.  This identity  holds for infinitely many $\beta \in \Bbb C$ if and only if  all $c_j = \overline c_j$; i.e.~all $c_j \in \Bbb R$.  So we see that $V(\beta)$ is constant along all arcs that do not contain any $\beta_j$ if and only if all $c_j \in \Bbb R$.

Relaxing the condition that $b_j \ne 0$ for all $j$ in Theorem 3 can be split into three mutually exclusive  cases:  (i) all $c_j \in \Bbb R$; (ii) all $c_j \in {\Bbb R} i$, and some $c_j = 0$; (iii)  some $c_j  \in {\Bbb R}-\{0\}$, and not all $c_j \in \Bbb R$.  Case (i) is covered in the next section.  Case (ii) requires a minor modification of the proof in Theorem 3, but the result still holds; if $c_j = 0$, then there may be finitely many trajectories $w(t)$ converging to $\beta_j$, which would correspond to finitely many trajectories on $\widetilde{Gp}$ converging to \nnmo\ states with the $j$th oscillator out of phase.

Case (iii) is more interesting.  Since not all $c_j \in \Bbb R$, the function  $V(\beta)$ is not constant on the circle, so most level curves of $V$ will meet the circle transversely.   Hence there will be an open set in $\Delta$ of initial conditions $w$ that will have $\Omega_+(w) =\{ \beta \}$ with $\beta \in S^1$, $\beta \ne \beta_j$.  The corresponding trajectories in $\widetilde{Gp}$ converge to sync.  We also see from the expansion in (\ref{Veq}) that if $c_j = a_j$ is real and nonzero, then we can find an open set of initial conditions $w \in \Delta$ whose trajectories $w(t)$ will converge to $\beta_j$ at all possible angles; the corresponding trajectories will converge to all possible \nnmo\ states with the $j$th oscillator out of phase.
Therefore in the full reduced state space $T^{N-1}$, in forward or reverse time, sync will attract a set of positive measure, but there will also be a  positive measure set of initial conditions which converge to a one-parameter family of neutrally stable \nnmo\ states.

\begin{figure}[h]
\centerline{\includegraphics[width=6.2in]{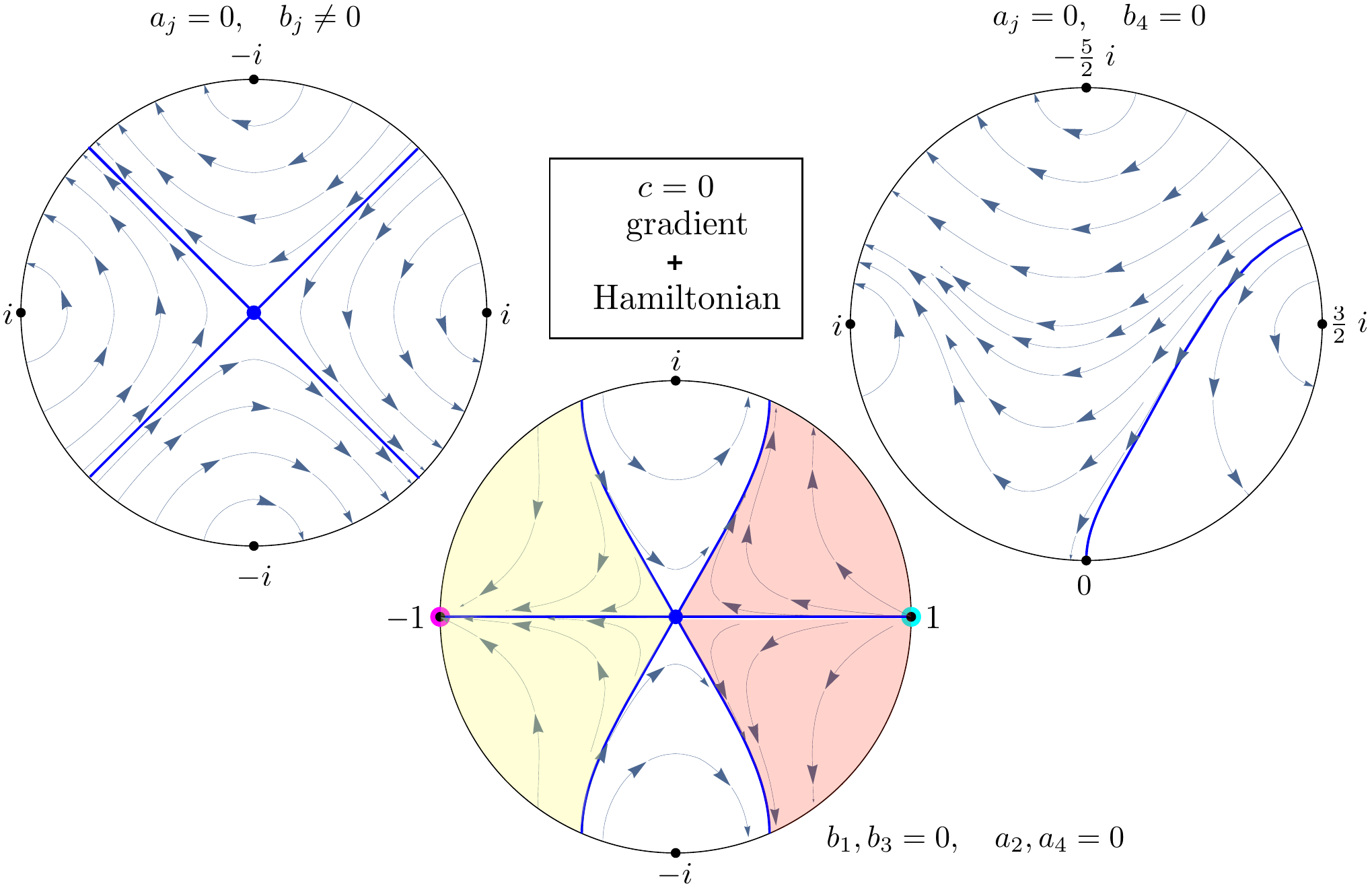}}
\caption{\label{fig4}
Three gradient $+$ Hamiltonian flows for N=4.
The values of the $c_j$ are indicated next to the corresponding $\beta_j$.
Trajectories with white background 
correspond to trajectories in \Tnmo\ that are homoclinic connections
to/from sync.
Trajectories with 
yellow (red) background
have
forward (backward) limits corresponding to \nnmo\ states.
}
\end{figure}

Figure~\ref{fig4} shows
three examples of gradient $+$ Hamiltonian flows.
We set $\beta_j=1,i,-1,-i$;
the values of the $c_j$ are indicated next to the corresponding $\beta_j$.
Trajectories with white background 
approach some point on the boundary circle distinct from the $\beta_j$ 
in forward and backward time;
the corresponding trajectories in \Tnmo\ are homoclinic connections
to/from sync.
In the top left panel 
the conditions of Theorem 3 apply:
almost all trajectories are homoclinic connections to and from sync.  There 
is one saddle point and
four heteroclinic connections between sync and the saddle indicated in blue.
The top right panel shows an example of case (ii) above where all the $c_i$ are imaginary but 
one is zero.  In this case there is a heteroclinic connection from sync to an \nnmo\ state
indicted by the blue trajectory.
The bottom panel 
has $c_j=\beta_j$
which is an example of case (iii).
In addition to homoclinic connections to and from sync (white background),
trajectories with yellow (red) background correspond to
heteroclinic connections from (to) sync to (from) 
an \nnmo\ state
The fixed point at $w=0$ is non-hyperbolic with 6 saddle connections.

\section{Special Case $\bf c_j \in {\Bbb R}, c = 0$; Connections To  2D Electrostatics}

We next turn to the even more special case with all $b_j = 0$ and $c = a = 0$; we assume that some $a_j \ne 0$ to avoid the trivial case. The associated $\dot w$ flow on the disc is gradient $+$ Hamiltonian, with holomorphic potential
$$
F(w) = 2\sum_{j = 1}^N a_j \log (w - \beta_j).
$$
The real part of $F(w)$  can be interpreted as the electrostatic potential for the 2D electric field on the plane given by point charges at $\beta_j$ with charge $a_j$ (up to some constant of proportionality depending on units, of course).  The $\dot w$  trajectories in $\Delta$ lie on the field lines for this electric field.  Let us call the oscillators {\sl positive, negative} or {\sl null} depending on whether the coefficients $a_j$ are positive, negative or zero respectively. 

\proclaim{Theorem 4.}{Consider the  K-S system with $c_j = a_j \in \Bbb R $ and $a=0$; assume some $ a_j \ne 0$.   Then almost all trajectories in the reduced state space $T^{N-1}$ are heteroclinic connections from an \nnmo\ state with a positive oscillator out of phase to an \nnmo\ state with a negative oscillator out of phase.}

\pf 
Assume $N \ge 3$ (the case $N=2$ is trivial; (\ref{BDyn}\ shows that the flow is identically $0$).  The trajectories for the $\dot w$ flow lie along the level sets of the function
$$
V(w) =  \Im F(w) = 2\sum_{j = 1}^N  a_j \arg (w - \beta_j).
$$
The $\dot w$ flow can have fixed points in the disc $\Delta$, which we saw above must be (perhaps higher order) saddles.  So we may have finitely many trajectories converging to a saddle point in $\Delta$.
In the previous section we showed that $V$ is constant along the arcs of $S^1 - \{ \beta_1,  \dots, \beta_N\}$.  Since $F$ is holomorphic near $\beta \ne \beta_j$, the level sets of the function $V = \Im F$ will have a single, smooth branch at any $\beta  \in S^1 - \{ \beta_1,  \dots, \beta_N\}$, unless $\beta$ is a critical point of $F$. This implies that a trajectory $w(t)$ for the $\dot w$ flow cannot converge to a point $\beta \in S^1$, with $\beta \ne \beta_j$, unless $\beta$ is a critical point for $F$, because the level set at $\beta$ is the arc of the circle containing $\beta$.     If $\beta$ is a critical point of $F$, the level sets of $V$ will have finitely many smooth branches intersecting in distinct angles at $\beta$, so we may have finitely many trajectories converging to $\beta$ in this case.  The same argument shows that there are at most finitely many trajectories converging to a null $\beta_j$.

Therefore all but finitely many trajectories $w(t)$ converge in forward and backward time to some non-null $\beta_j$.  The real potential for this flow is
$$
U(w) = \sum_{j = 1}^N a_j \log |w- \beta_j|,
$$
and $\dot U > 0$ along non-trivial trajectories.  If $w(t)$ converges to a non-null $\beta_j$ in forward time, then we must have $a_j < 0$; otherwise $U(w(t)) \to  - \infty$. Therefore all but finitely trajectories converge to negative (resp.~positive)  $\beta_j$ in forward (resp.~backward) time (the ``test charge'' for the flow is positive). The trajectories converging to each non-null $\beta_j$ are in one-to-one correspondence to all asymptotic angles of approach to $\beta_j$, because $V(w(t))$ is conserved along trajectories.  So we see that there is a one-parameter family of trajectories converging in forward time to the negative $\beta_j$, and a one-parameter family converging in backward time to the positive $\beta_j$.  This gives a complete picture of the dynamics of the $\dot w$ flow.  The corresponding trajectories in the reduced group orbit $\widetilde{Gp}$ converge in forward (resp.~backward) time to all possible \nnmo\ states with a negative (resp.~positive) oscillator out of phase.  The finitely many exceptional trajectories converge to an asynchronous fixed state in $\widetilde{Gp}$, or to an \nnmo\ state with a null oscillator out of phase, or to sync, in forward or backward time.

\qed

\begin{figure}[h]
\centerline{\includegraphics[width=5.2in]{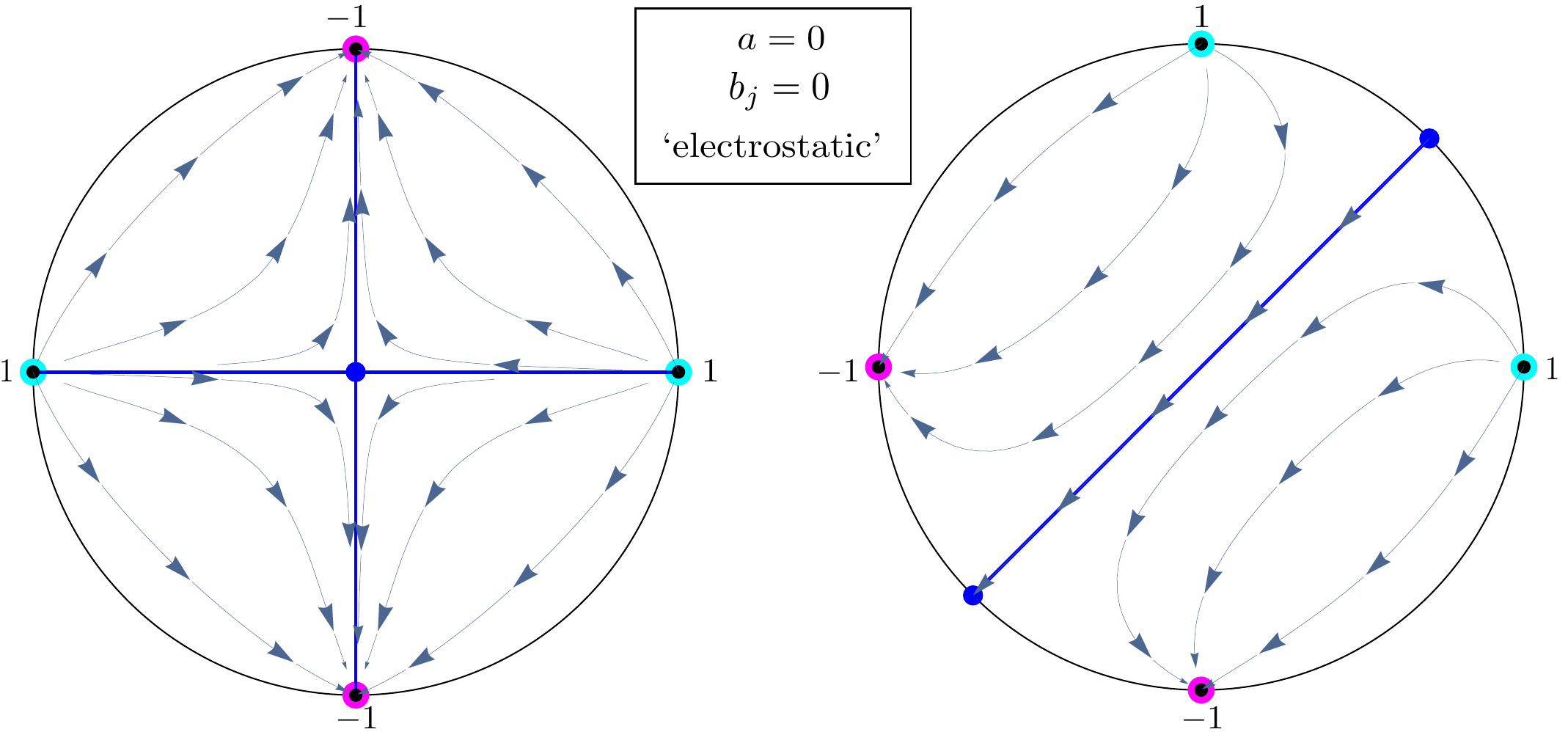}}
\caption{\label{fig5}
Two gradient $+$ Hamiltonian $\dot w$ flows with real $c_j$ for $N=4$
(``electrostatic'' case).
In both examples two oscillators have ``charge'' $+1$, two have ``charge'' $-1$.
}
\end{figure}

Figure~\ref{fig5} shows two examples of phase portraits in the electrostatic case for $N=4$.
In both cases
two of the oscillators have ``charge'' $+1$ and two have ``charge'' $-1$.
In the first panel the charges alternate around the circle, and there is a saddle at $w=0$.
In the second panel 
the electric field has 
two saddles 
at $\pm e^{\pi i /4}$ 
on the boundary circle
(blue dots)
with
an exceptional trajectory 
connecting them,
which corresponds to a homoclinic orbit to and from sync in $T^{N-1}$.
Note that the 
$c_j$ for the 1st panels of Figures \ref{fig4}~and \ref{fig5}~are related by a factor of
$i$, so these two $\dot w$ flows are orthogonal.
Both flows are simultaneously gradient and Hamiltonian 
but with functions $U(w)$ and $V(w)$ exchanged.  
We conclude this section by mentioning that in the course of our work above, we proved that if finitely many point charges are placed on a circle, and the total charge is $0$, then the corresponding $2D$ electric field is everywhere tangent to the circle.  This is clearly false if the total charge is not $0$.

\section{Case $\bf c_j > 0$: Hyperbolic Barycenters And The Symmetric K-S Model}

Our final special case is when all $c_j =a_j > 0$, which includes the original symmetric Kuramoto model with identical $c_j = K/N > 0$.  Note that the results in Theorem 1 apply in this case.  However as we shall see below, under the additional assumption $a_j >0$ there is at most one fixed point for the $\dot w$ flow, this fixed point is always repelling, and the two possible generic asymptotic behaviors in backward time in Theorem 1 are mutually exclusive in this special case.

\proclaim{Theorem 5.}{Consider the  K-S system with $c_j  = a_j > 0 $ for all $j$.   Then almost all trajectories in the reduced state space $T^{N-1}$ converge in forward time  to sync.  If all $a_j < a/2$ then almost all trajectories  converge in backward time to a fully asynchronous state; if some $a_j \ge a/2$, then almost all trajectories converge in backward time to the \nnmo-state with the $j$th coordinate out of phase by $\pi$. }

\pf 
Assume $N \ge 3$ (the case $N=2$ follows easily from (\ref{BDyn}).   Choose any base point $p$ whose coordinates $\beta_j$ are all distinct.  The forward time dynamics follow from Theorem 1.   In the proof of Theorem 1, we saw that in backward time all trajectories $w(t)$ for (\ref{weq}) converge either to a fixed point $w^\ast \in \Delta$ or to some $\beta_j$.  The corresponding polar coordinate $\theta(t)$ must converge to $\theta_j^\ast$, which is $\pi/2$ since the $c_j$ are real.  The linearized $\dot r$ equation in (\ref{eqrlin}) simplifies to $\dot r = r(2a_j - a)$; this shows that we cannot have $r(t) \to 0$, $\theta(t) \to \pi/2$  as $t \to -\infty$ if $a_j < a/2$.   So if all $a_j < a/2$, then all trajectories $w(t)$ must converge in backward time to a fixed point $w^\ast$, and the corresponding trajectories converge in backward time to a fully asynchronous state in $\widetilde{Gp}$.

On the other hand, if some $a_j \ge a/2$, then the $\dot w$ flow (\ref{weq}) has no fixed points; the expression
$$
\sum_{k = 1}^N a_j  M_w(\beta_k)
$$
is a positive weighted average of $N \ge 3$ distinct points $M_w (\beta_k)$ on the unit circle, with weights summing to $a$ and one weight at least $a/2$; this can never be $0$.  Therefore all trajectories must converge in backward time to the unique $\beta_j$ which has $a_j \ge a/2$, and the corresponding trajectories converge in backward time to the \nnmo\ state which has the $j$th oscillator out of phase by $\pi$.
\qed

We can say a bit more about the $\dot w$ dynamics in this case.  Suppose $N \ge 3$, and all $a_j < a/2$.  The eigenvalues at a fixed point $w^\ast$ from (\ref{ev}) are 
$$
\lambda_\pm = {a \over 2} \left (1  \pm \Bigl | \sum_{j=1}^N {a_j \over a} M_{w^\ast} (\beta_j)^2 \Bigr |\right).
$$
The sum in the equation above is a weighted average of $N \ge 3$ points on the unit circle, which are not all identical, and hence must have absolute value less than $1$.  Hence $\lambda_\pm > 0$, so all fixed points are repelling.   All trajectories $w(t)$ converge in backward time to a fixed point, so we must have at least one fixed point.  If there were more than one fixed point, then the backward time dynamics would partition the disc into two or more basins of attraction of the fixed points, which are disjoint open sets, contradicting the connectedness of the disc.  
(We gave a somewhat different proof of the existence and uniqueness of fixed points 
in~\cite{chen2017hyperbolic}).

The unique fixed point $w^\ast$ is known as the {\sl hyperbolic barycenter} of the configuration $\beta_j$ of points on the circle with weights $a_j$;  it is the unique point at which the weighted sum of the unit vectors pointing at all the $\beta_j$ is $0$.  In backward time, all trajectories $w(t)$ converge to the barycenter $w^\ast$.  In forward time, there is a unique trajectory converging to each point $\beta \in S^1$; if $\beta \ne \beta_j$, the corresponding trajectory in $\widetilde{Gp}$ converges to sync; if $\beta = \beta_j$, the corresponding trajectory is the saddle trajectory converging to the \nnmo\ state with the $j$th oscillator out of phase by $\pi$.
If some $a_j \ge a/2$, then the $\dot w$ flow has no fixed points, and in backward time all trajectories converge to $\beta_j$.  The forward time dynamics are the same as in the previous case, except that no trajectory converges to  $\beta_j$ in forward time.  

\begin{figure}[h]
\centerline{\includegraphics[width=5.2in]{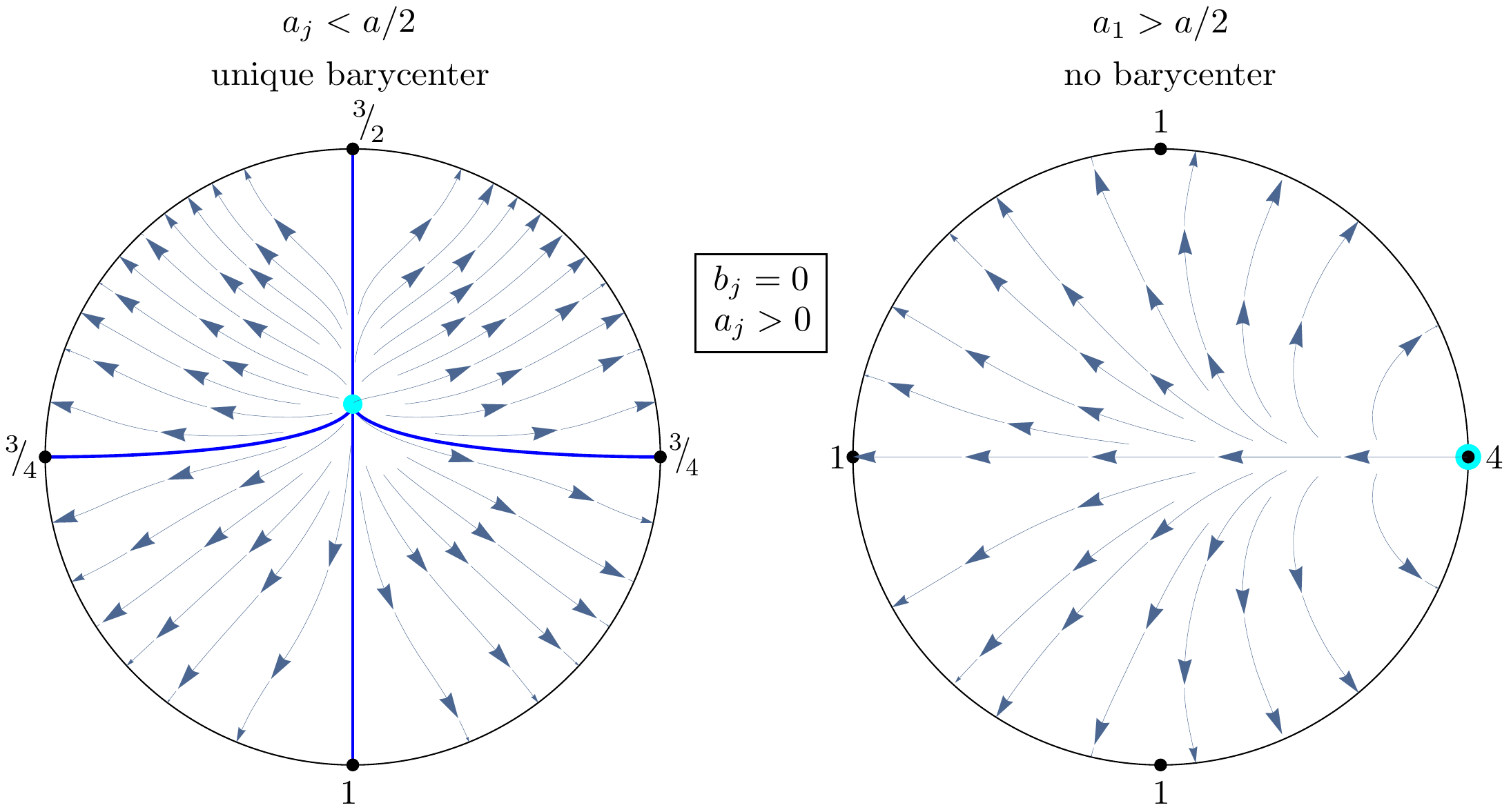}}
\caption{\label{fig6}
First 
example has $c_j=a_j>0$ and $a_j<a/2$;
there is a unique fixed point at the hyperbolic barycenter.
Second example has $c_j=a_j>0$ but $a_1>a/2$;
there is no fixed point in the disc, and 
the \nnmo\ state in \Tnmo\ with $\theta_1$ out of phase by $\pi$
is repelling.
}
\end{figure}

Figure \ref{fig6} shows two examples of phase portraits 
in the case
$c_j=a_j>0$.
In the first panel 
all $a_j<a/2$;
there is a unique, repelling fixed point at the hyperbolic barycenter of the $\beta_j$.
In the second panel 
$a_1>a/2$;
there are no fixed points in the disc,
and 
all trajectories converge in backward time to $\beta_1=1$ at angle $\pi/2$,
which corresponds to a repelling \nnmo\ state with $\theta_1$ 
out of phase by $\pi$.

We remark that the condition $a_j > 0$ can be relaxed to $a_j \ge 0$ as long as not all $a_j = 0$;
the arguments in Theorem 5 go through, with one exceptional case: if exactly two of the $a_j$, say $a_1$ and $a_2$,  are nonzero and equal, and all other $a_j = 0$.  
Then (\ref{Req}) shows that $\dot r = 0$, and $\theta(t) \to \pi/2$ as $t \to -\infty$.  
So all trajectories $w(t)$ converge in backward time  to  some point on the line of fixed points, which is the geodesic arc joining $\beta_1$ and $\beta_2$.  The corresponding trajectories in $T^{N-1}$ converge in backward time to asynchronous states.  In forward time almost all trajectories go to sync as before.

\pagebreak

\section{Discussion}

Using M\"obius group and hyperbolic geometry techniques, we are able to give a fairly complete description of the long-term dynamics of the asymmetric Kuramoto-Sakaguchi network (\ref{AKS}) on the reduced state space $T^{N-1}$.
A summary of our findings is given in the following table.  In each case, the description of the dynamics as $t \to \pm \infty$ is the generic behavior; there may be a set of measure zero of exceptional trajectories with different behavior.

\begin{table}[h]
\begin{center}
\begin{tabular}{|l|l|l|}
\hline
 Case & Dynamics as $t \to \infty$ \ \  & Dynamics as $t \to - \infty$ 
\\ 
\hline
\hline
1. 
$a > 0$ & sync & asynchronous states
\\
        &      & finitely many $(N\!-\!1,1)$ states
\\
\hline
\hline
2. 
$c = ib \ne 0$,        & sync & same as $t \to \infty$
\\
$\hphantom{2.}$~
all $b_j \ne b/2$      & periodic orbits   & 
\\
\hline
3. 
$c=ib \ne 0$,          & sync & same as $t \to \infty$
\\
$\hphantom{3.}$~
some $b_j = b/2$       & periodic orbits & 
\\
                       & $(N\!-\!1,1)$ states & 
\\
\hline
\hline
4. 
$c=0$,  all $b_j \ne 0$ & sync & same as $t \to \infty$
\\
\hline
5.  
$c=0$, all $b_j=0$,     & \nnmo\ states & same as $t \to \infty$
\\
$\hphantom{5.}$~
some $a_j \ne 0$        & & 
\\
\hline
\hline
6.  
All $c_j =a_j > 0$,     & sync & asynchronous states
\\
$\hphantom{6.}$~
$a_j < a/2$             & & 
\\
\hline
7.  
All $c_j =a_j> 0$,       & sync & $(N\!-\!1,1)$ state with $j$th 
\\
$\hphantom{7.}$~
some $a_j \ge a/2$       &      & oscillator $\pi$ out of phase
\\
\hline
\end{tabular}
\begin{tabular}{l}
 \\ \\ \\ 
\\
\multirow{5}{*}{\Bigg \}\ \ Hamiltonian}
\\
\\
\\
\\
\\
\multirow{3}{*}{\Bigg \}\ \ gradient $+$ Hamiltonian}
\\ 
\\
\\
\\
\multirow{3}{*}{\Bigg \}\ \ gradient}
\\
\\
\\
\\
\end{tabular}
\end{center}
\caption{\label{tab1}
Summary of the generic dynamics for the system (\ref{KsysC})
on the reduced state space \Tnmo
as $t\to\pm\infty$. 
A measure zero set of initial conditions may have exceptional behavior.
}
\end{table}

The hyperbolic-geometric approach we developed
facilitates the analysis 
of the somewhat subtle dynamics near the \nnmo\ states.
For example, in the case of the symmetric Kuramoto model with $a>0$,
our boundary flow analysis shows that there is a codimension one
set of initial conditions that flow to \nnmo\ saddles.
This point is omitted in the classic 
paper~\cite{watanabe1994constants}.
This framework also supports the analysis of the system dynamics
in the Hamiltonian and other special cases,
where the asymmetric model can exhibit 
more complex dynamics than the symmetric model.
For example, this analysis gives the existence of
homoclinic and heteroclinic
non-periodic orbits 
to/from sync and \nnmo\ states
in the Hamiltonian case.

We expect that the methods we developed in this paper will have further applications
to Kuramoto networks with higher-order order parameters, 
or Kuramoto networks consisting of two or more populations of oscillators with different natural 
frequencies.  
Additionally, the possibility in the $c_j$ model of prescribing the location and 
stability of any number of competing fixed points 
may prove useful in applications to machine learning or reservoir computing.
We hope to take up this study in the future.  
Another interesting direction in which to extend this framework is to consider
networks of 
``oscillators'' where the state space for an individual oscillator
is a manifold other than $S^1$, 
such as spheres of dimension 2 or greater;
see for example
\cite{lohe2009non,lohe2018higher,jacimovic2018low}.
Finally, we thank our colleagues and friends Martin Bridgman, Kathryn Lindsey, Curtis McMullen and Robert Meyerhoff, all experts on hyperbolic geometry and complex dynamics, for many conversations and comments that were not just helpful, but indispensable in bringing this project to a conclusion.

This work was supported by NSF Grant DMS 1413020.

\bibliography{refs_Jan2017}



\end{document}